\documentclass[]{jfm_old}

\usepackage{graphicx} \graphicspath{ {./figure} }
\usepackage{epstopdf,epsfig}
\usepackage{newtxtext}
\usepackage{newtxmath}
\usepackage{natbib}

\newcommand{\RomanNumeralCaps}[1]
\linenumbers

\usepackage{amssymb}
\usepackage{latexsym}
\usepackage{amsmath,mathtools}
\numberwithin{equation}{section}
\usepackage{bm}
\usepackage{multirow}
\usepackage{arydshln}
\usepackage{lipsum, xcolor}

\title{Transient growth of a wake vortex and its initiation via inertial particles}

\shortauthor{S. Lee and P. S. Marcus}
\shorttitle{Transient growth of a wake vortex and its initiation via inertial particles}

\author{Sangjoon Lee\aff{1}
 \and Philip S. Marcus\aff{1}
 \corresp{\email{pmarcus@me.berkeley.edu}}}

\affiliation{\aff{1}Department of Mechanical Engineering, University of California,
Berkeley, CA 94720, USA}

\begin{document}
\maketitle

\begin{abstract}
The transient dynamics of a wake vortex, modelled as a strong swirling $q$-vortex, are investigated with a focus on optimal transient growth driven by continuous eigenmodes associated with continuous spectra. The pivotal contribution of viscous critical-layer eigenmodes \citep[][\textit{J. Fluid Mech.}, vol. 967]{Lee2023} amongst the entire eigenmode families to optimal perturbations is numerically confirmed, utilising a spectral collocation method for a radially unbounded domain that ensures correct analyticity and far-field behaviour. The consistency of the numerical method across different sensitivity tests supports the reliability of the results and provides flexibility for tuning. Both axisymmetric and helical perturbations with axial wavenumbers of order unity or less are examined through linearised theory and non-linear simulations, yielding results that align with existing literature on energy growth curves and optimal perturbation structures. The initiation process of transient growth is also explored, highlighting its practical relevance. Inspired by ice crystals in contrails, the backward influence of inertial particles on the vortex flow, particularly through particle drag, is emphasised. In the pursuit of optimal transient growth, particles are initially distributed at the periphery of the vortex core to disturb the flow. Two-way coupled vortex-particle simulations reveal clear evidence of optimal transient growth during ongoing vortex-particle interactions, reinforcing the robustness and significance of transient growth in the original non-linear vortex system over finite time periods.
\end{abstract}


\section{Introduction}\label{sec:introduction}
Wake vortex following an aircraft is widely recognised for its long-lived presence over time, which has made it a significant focus of aerodynamic research. It holds importance, particularly in comprehending the mechanisms behind its decay process. Rapid destruction of the wake vortices is deemed beneficial in several aspects, including enhancing air traffic safety by mitigating wake-related hazards and improving airport operation efficiency by reducing intervals between aircraft during take-off and landing on the same runway \citep{Spalart1998, Hallock2018}. Additionally, wake vortices contribute to the development of condensation trails (or contrails), whose impact on climate change via radiative forcing has been actively assessed \citep[e.g.,][]{Schumann2005, Naiman2011, Lee2021}, by capturing the jet exhaust particles around the low-pressure vortex core, facilitating the formation of ice crystals. The early demise of wake vortices may impede early contrail development and, as a result, potentially influence its subsequent climate impact.

There are several factors influencing the decay process of the wake vortex, including stratification \citep{Sarpkaya1983}, ground effect \citep{Proctor2000}, and various surrounding conditions \citep[see][p. 30]{Hallock2018}. Among these, the activation of wake vortex instability generally provides the most effective pathway for vortex breakup. Classical wake vortex instability mechanisms have been studied in the context of the typical counter-rotating vortex pair configuration for aircraft trailing vortices \citep{Crow1970, Moore1975, Tsai1976}, where one vortex is disturbed by the strain induced by the other. If an infinitesimal perturbation (or eigenmode) of the base vortex profile exhibits a positive real growth rate (or eigenvalue), it can be triggered by atmospheric turbulence \citep[e.g.,][]{Crow1976} and grow exponentially until nonlinear dynamics take over, ultimately leading to the linkage of the two vortices. In this context, growth is evidenced by the presence of an unstable eigenmode, which serves as a solution to the Navier-Stokes or Euler equations linearised around the base vortex profile, a process commonly referred to as linear instability analysis.

However, for an undisturbed wake vortex — such as one not influenced by nearby vortices and typically modelled as the Batchelor vortex \citep{Batchelor1964} — linear instability analysis generally shows that a strong swirling vortex remains stable. In most inviscid cases, the vortex is linearly neutrally stable unless accompanied by a strong axial velocity component \citep{Stewartson1983, Stewartson1985, Heaton2007Centre}. Several experiments suggest that the axial velocity of realistic wake vortices is generally not strong enough to make the system linearly unstable \citep[e.g.,][]{Leibovich1978, Fabre2004}. With the inclusion of viscosity, \citet{Fabre2004} demonstrated that centre-mode instabilities can occur even with moderate axial velocity, where the instability is primarily concentrated near the vortex core. However, this instability remains weak \citep[][p. 496]{Heaton2007Optimal}, and its practical relevance is uncertain. In general, viscosity has been observed to exert a predominantly stabilising effect on eigenmodes \citep[see][p. 198]{Khorrami1991}.

To unravel the early development of a single vortex, several approaches have been employed with varying degrees of success. One such approach is the analysis of resonant triad instability (RTI), which examines instability arising from the resonance of two secondary modes, induced by the primary mode acting as a disturbance to the vortex \citep[e.g.,][]{Mahalov1993, Wang2024}. In the context of vortex stability, the RTI mechanism represents a generalised version of the elliptical instability \citep{Moore1975, Tsai1976}, where the primary disturbance is the strain generated by a neighbouring vortex.

Another approach is transient growth analysis, which investigates an optimal initial perturbation (typically represented as a sum of eigenmodes) that can exhibit significant energy growth over finite times, even as it decays asymptotically as time approaches infinity \citep[e.g.,][]{Schmid1994, Heaton2007Optimal, Mao2012, Navrose2018}. This behaviour results from the non-normality of the linearised Euler or Navier-Stokes operators, producing families of continuously varying eigenmodes \citep{Mao2011, Lee2023}, in addition to discrete ones. Several studies have applied transient growth analysis to vortices with axial flows \citep{HeatonPeake, Mao2012} and, similarly, to jets with swirling flows \citep{Muthiah2018}, highlighting the significance of continuous eigenmodes.

In this paper, we investigate transient growth in wake vortices under physically relevant conditions (i.e., with non-zero viscosity), building on earlier studies that examined the role of continuous eigenmodes in driving transiently growing perturbations. In the presence of viscosity, continuous eigenmodes can be categorised into multiple distinct families. This leads to an important question: Which family contributes most significantly to optimal perturbations for transient growth—the potential family \citep{Mao2011}, the viscous critical-layer family \citep{Lee2023}, or both? This study primarily aims to identify the dominant eigenmode family, enabling a more focused analysis of optimal transient growth by excluding less influential families. For readers seeking further information on potential and viscous critical-layer eigenmodes, discussions are presented in \S \ref{sec:numericalsensitivity}, along with an illustration in figure \ref{fig:pot_vs_visc}; additional details can be found in \citet{Lee2023}.  

Subsequently, another crucial aspect to consider is that the numerically resolvable portion of continuous eigenmodes depends on the discretisation scheme of the method used. Addressing the aforementioned question also provides insights into the appropriate methodology for tackling this type of problem. In the literature, Chebyshev spectral collocation methods have been commonly employed for wake vortex stability analysis \citep[e.g.,][pp. 354-357]{Ash1995}. In contrast, \citet{Lee2023} proposed the mapped Legendre spectral collocation method, successfully distinguishing the viscous critical-layer family for the first time, despite its resemblance to the potential family. In this study, we extend the application of the mapped Legendre spectral collocation method to the transient growth analysis of wake vortices, highlighting its effectiveness, particularly for radially unbounded swirling flow problems.

Finally, we consider ice particles as a potential source for initiating optimal perturbations during the early stages of vortex development, leading to transient growth. Computational fluid dynamics (CFD) studies have explored the interaction between jet exhaust and vortices in the context of contrail formation, typically involving ice microphysics \citep[][]{Lewellen2001, Shirgaonkar2007, Paoli2005, Naiman2011}. However, to the best of our knowledge, the role of drag momentum exchange from jet exhaust (or ice particles) in influencing short-term wake vortex development remains unclear, despite its potential importance. We anticipate that particle drag can significantly displace the vortex over a short period if particles cluster around the vortex core during jet entrainment, triggering temporarily large-growing perturbations at the vortex core's periphery \citep[e.g., ][p. 43]{Mao2012}. In the early stages of jet exhaust, individual particles can grow to only a few microns, but total particle number density is reported to be high ($10^{9}$ to $10^{11}$ per cubic metre) \citep[]{Paoli2004, Paoli2005}, making their bulk effect on momentum exchange non-negligible.
This study is the first step to investigate the role of particle concentration near the vortex in the initiation of transient growth.

The remainder of the paper is structured as follows. In \S 2, the essence of the linear stability analysis of wake vortices \citep{Lee2023} is revisited and then incorporated into a transient growth analysis. In \S 3, the optimal perturbation structures obtained from this analysis are presented, identifying the continuous eigenmode family that makes the dominant contribution. In \S 4, the initiation of optimal perturbations via inertial particles near the vortex is examined. In \S 5, the overall findings are summarised and concluded.

\section{Transient growth formalism}\label{sec:transientgrowthformalism}
\subsection{Formulation}\label{sec:formulation}
We briefly revisit the gist of the linear stability analysis of wake vortices by \citet{Lee2023} and then incorporate it into a transient growth analysis. Unless specified otherwise, we use a cylindrical coordinate system $(r, \phi, z)$. All variables are non-dimensionalised with respect to the characteristic radial length scale, $R_0$, the characteristic azimuthal velocity scale, $U_0$, and the fluid density $\rho$. Detailed definitions of $R_0$ and $U_0$ can be found in \citet[p. 755]{Lessen1974}. The base velocity profile $\overline{\bm{U}}$, represented as the Batchelor vortex \citep{Batchelor1964}, or the $q$-vortex in its non-dimensional form, is given by
\begin{equation}
    \overline{\bm{U}}(r) = \left(\frac{1-e^{-r^2}}{r}\right) \hat{\bm{e}}_\phi + \left(\frac{1}{q}e^{-r^2}\right) \hat{\bm{e}}_z ,
    \label{qvortdimless}
\end{equation}
where $q$ is the swirl parameter that determines the relative strength of the swirling motion. The vortex core region is defined as the radial location where the azimuthal velocity component is maximised, which is $r \le 1.12$.

The governing equations of fluid motion assume a Newtonian fluid with constant density, $\rho$, and constant kinematic viscosity, $\nu$. In terms of the total velocity, $\bm{u} \equiv u_r \hat{\bm{e}}_r + u_\phi \hat{\bm{e}}_\phi + u_z \hat{\bm{e}}_z$, and the total specific energy, $\varphi \equiv (\bm{u} \cdot \bm{u}) / 2 + p$, where $p$ denotes the total pressure (non-dimensionalised by $\rho U_0^2$), they are expressed as
\begin{equation}
    \frac{\partial \bm{u}}{\partial t} = - \bm{\nabla} \varphi + \bm{u} \times \bm{\omega}  + \frac{1}{\Rey} {\nabla}^2 \bm{u}~~~\text{with}~\bm{\nabla} \cdot \bm{u} = 0,
    \label{nonlingoveqn}
\end{equation}
where $\bm{\omega} \equiv \bm{\nabla} \times \bm{u}$ is the total vorticity, and $\Rey \equiv U_{0}R_{0} / \nu$ is the Reynolds number. The equations are linearised around the base flow profile by decomposing $\bm{u}$ and $p$ into base terms (indicated by overbars $\overline{\ast}$) and perturbations (indicated by primes $\ast'$). Using the toroidal-poloidal decomposition with $\hat{\bm{e}}_z$ as a reference vector, the resulting form becomes
\begin{equation}
    \frac{\partial }{\partial t} \begin{pmatrix}\psi' \\ \chi' \end{pmatrix} = \mathbb{P} \big( \bm{\overline{U}}(r) \times \bm{\omega}' \big) - \mathbb{P} \big( \bm{\overline{\omega}}(r) \times \bm{u}' \big) + \frac{1}{\Rey} {\nabla}^2 \begin{pmatrix}\psi' \\ \chi' \end{pmatrix},
    \label{lingoveqn}
\end{equation}
where $\psi' (r,\phi,z,t)$ and $\chi' (r,\phi,z,t)$ are the toroidal and poloidal streamfunctions of $\bm{u}' (r,\phi,z,t)$, respectively. The operator $\mathbb{P}$ decomposes a smooth vector field into its toroidal and poloidal scalar components. If the input vector field is solenoidal, $\mathbb{P}$ is invertible; in other words, if $\bm{\nabla} \cdot \bm{u}' = 0$ and $\mathbb{P} ( \bm{u}' ) = (\psi',~\chi')$, then $\bm{u}'$ can be uniquely reconstructed from $(\psi',~\chi')$ \citep[for further details, see][pp. 9-11]{Lee2023}. Thus, \eqref{lingoveqn} only involve two state variables: $\psi'$ and $\chi'$. These reduced form facilitates the imposition of the analyticity constraint at $r=0$, as each state variable is treated independently of one another without coupling.

If we introduce the Fourier ansatz (indicated with tildes $\Tilde{\ast}$) for the azimuthal and axial wavenumbers $m \in \mathbb{Z}$, $\kappa \in \mathbb{R}\setminus \left\{ 0 \right\}$, to represent perturbations of finite axial wavelengths, i.e.,
\begin{equation}
    \begin{pmatrix}\psi' \\ \chi' \end{pmatrix} = \begin{pmatrix} \Tilde{\psi}(r,t) \\ \Tilde{\chi}(r,t) \end{pmatrix} e^{\mathrm{i}(m\phi + \kappa z)} ,
    \label{ansatz}
\end{equation}
\eqref{lingoveqn} further reduces to a spatially one-dimensional form expressed as
\begin{equation}
    \frac{\partial \Tilde{\bm{\wp}}}{\partial t} = \mathcal{L}_{m \kappa}^{\nu} ( \Tilde{\bm{\wp}} ),
    \label{lingoveqn-simp}
\end{equation}
where $\Tilde{\bm{\wp}} (r,t) \equiv ( \Tilde{\psi} (r,t),~\Tilde{\chi} (r,t) )$ represents the toroidal-poloidal streamfunction set (equivalent to its corresponding velocity Fourier ansatz, $\Tilde{\bm{u}} (r,t) e^{\mathrm{i}(m\phi + \kappa z)}$). Here, $\mathcal{L}_{m\kappa}^{\nu}$ is the linear operator with respect to $\Tilde{\bm{\wp}}$, representing the right-hand side of \eqref{lingoveqn} with the inclusion of \eqref{ansatz}. Note that the operator varies with the wavenumbers $m$ and $\kappa$ and the Reynolds number $\Rey$, as indicated by the subscript and superscript.

To obtain physically meaningful solutions to \eqref{lingoveqn-simp} in an unbounded domain ($0 \le r < \infty$), the analyticity at the origin and rapid decay as $r \rightarrow \infty$ are necessary. The most prevalent discretisation schemes for computing these solutions have been Chebyshev spectral collocation methods \citep[e.g.,][]{Ash1995, Antkowiak2004, Fontane2008, Mao2011, Muthiah2018}, which uses a bounded domain requiring two closed ends and therefore demands approximations of the above constraints. In contrast, the mapped Legendre spectral collocation method, described by \citet{Lee2023}, is specifically designed for unbounded domains while accurately satisfying these conditions without additional treatments. When the problem is discretised using either method, we obtain
\begin{equation}
    \frac{d \Tilde{\bm{\varrho}}}{d t} = \mathsfbi{L}_{m\kappa}^{\nu} \Tilde{\bm{\varrho}} ,
    \label{ode-disc}
\end{equation}
where $\Tilde{\bm{\varrho}} (t)$ is the discretised version of $\Tilde{\bm{\wp}} (r,t)$ in spectral space, consisting of the spectral coefficients of $\Tilde{\psi}$ and $\Tilde{\chi}$ in order, and $\mathsfbi{L}_{m\kappa}^{\nu}$ is the matrix expression of $\mathcal{L}_{m\kappa}^{\nu}$.

Similarly, we define the discretised version of $\Tilde{\bm{u}} (r,t)$ in physical space as $\Tilde{\bm{\upsilon}} (t)$, which consists of the collocated values of $\Tilde{u}_r$, $\Tilde{u}_\phi$ and $\tilde{u}_z$ in order. The conversion between $\Tilde{\bm{\varrho}}$ and $\Tilde{\bm{\upsilon}}$ can be achieved through the matrix expression of $\mathbb{P}$, denoted as $\mathsfbi{P}$. The construction of $\mathbb{P}$ based on the mapped Legendre spectral collocation method is described in \citet[][pp. 330-333]{Matsushima1997}. We use the following notations: $\Tilde{\bm{\varrho}} = \mathsfbi{P}^{\dagger} \Tilde{\bm{\upsilon}}$ and $\Tilde{\bm{\upsilon}} = \mathsfbi{P} \Tilde{\bm{\varrho}}$. Under the solenoidal velocity assumption, both $\mathsfbi{P}^{\dagger}\mathsfbi{P}$ and $\mathsfbi{P}\mathsfbi{P}^{\dagger}$ can be treated as identity maps. Now we define the `energy' $E$ of a velocity of the form $\Tilde{\bm{u}} (r,t) e^{\mathrm{i}(m\phi + \kappa z)}$ as
\begin{equation}
    E(\Tilde{\bm{u}}) \equiv \int_{0}^{\infty} \left( \Tilde{u}_r^* \Tilde{u}_r + \Tilde{u}_\phi^* \Tilde{u}_\phi + \Tilde{u}_z^* \Tilde{u}_z  \right) rdr.
    \label{energy-integ}
\end{equation}
A similar usage can be found in \citet{Mao2012}. Using numerical integration, e.g., a quadrature rule, $E(\Tilde{\bm{u}})$ is expressed as
\begin{equation}
    E(\Tilde{\bm{u}}) = \Tilde{\bm{\upsilon}}^* \mathsfbi{M} \Tilde{\bm{\upsilon}} =
    \Tilde{\bm{\varrho}}^* \mathsfbi{P}^* \mathsfbi{M} \mathsfbi{P} \Tilde{\bm{\varrho}} ,
    \label{energy-matrix}
\end{equation}
where $\mathsfbi{M}$ represents the numerical integration form of \eqref{energy-integ}. A specific example of $\mathsfbi{M}$ using the Gauss-Legendre quadrature rule is given in Appendix \ref{appA}.

At last, we apply the transient growth formalism \citep{Schmid2001, Mao2012, Muthiah2018} in order to complete our formulation. Consider a set of eigenmodes of $\mathsfbi{L}_{m\kappa}^{\nu}$ containing $p$ elements $\left\{ \Tilde{\bm{\varrho}}_1,~\Tilde{\bm{\varrho}}_2, \cdots,~ \Tilde{\bm{\varrho}}_p \right\}$, corresponding to eigenvalues $\left\{ \sigma_1,~\sigma_2,\cdots,~\sigma_p \right\} \subset \mathbb{C}$, respectively. Assuming that $\Tilde{\bm{\varrho}}$ belongs to the eigenspace spanned by these $p$ eigenmodes, i.e.,
\begin{equation}
    \Tilde{\bm{\varrho}} (t) = \sum_{k=1}^{p} \Tilde{\xi}_k e^{\sigma_k t} \Tilde{\bm{\varrho}}_k,
    \label{changeofbasis}
\end{equation}
we use a new vector $\Tilde{\bm{\xi}}_0 \equiv (\Tilde{\xi}_1, \Tilde{\xi}_2, \cdots, \Tilde{\xi}_p ) \in \mathbb{C}^p$ to represent $\Tilde{\bm{\varrho}}$. For instance, at time $t=\tau$, $\Tilde{\bm{\varrho}} (\tau)$ is expressed as $\exp (\tau \mathsfbi{S}) \Tilde{\bm{\xi}}_0$, where $\mathsfbi{S} \equiv \text{diag}(\sigma_1, \sigma_2, \cdots, \sigma_p)$. Focusing on the transient growth process, the chosen eigenmodes are assumed to be asymptotically stable in time ($\Real (\sigma_k) < 0$), which mostly holds for strong swirling $q$-vortices. By defining $\mathsfbi{V} \equiv \left( \begin{array}{c;{2pt/1pt}c;{2pt/1pt}c;{2pt/1pt}c}\Tilde{\bm{\varrho}}_1 & \Tilde{\bm{\varrho}}_2 & \cdots & \Tilde{\bm{\varrho}}_p \end{array} \right)$, \eqref{changeofbasis} at time $t = \tau$ becomes
\begin{equation}
    \Tilde{\bm{\varrho}} (\tau) = \mathsfbi{V} {\exp (\tau \mathsfbi{S})} \Tilde{\bm{\xi}}_0,
    \label{changeofbasis_2}
\end{equation}
and applying this to \eqref{energy-matrix} leads to the energy formula using the following $\ell^2$ norm
\begin{equation}
\begin{aligned}
    E(\Tilde{\bm{u}}(\tau)) & = 
    \Tilde{\bm{\xi}}_0^* {\exp (\tau \mathsfbi{S}^*)} \mathsfbi{V}^* \mathsfbi{P}^* \mathsfbi{M} \mathsfbi{P}  \mathsfbi{V} {\exp (\tau \mathsfbi{S})} \Tilde{\bm{\xi}}_0 \\
    & = \big\lVert \mathsfbi{F} \exp (\tau \mathsfbi{S}) \Tilde{\bm{\xi}}_0 \big\rVert_2^2,
    \label{energy-l2norm}
\end{aligned}
\end{equation}
where the matrix $\mathsfbi{F}$ is defined such that $\mathsfbi{F}^* \mathsfbi{F} = \mathsfbi{V}^* \mathsfbi{P}^* \mathsfbi{M} \mathsfbi{P} \mathsfbi{V}$. The maximum energy growth, $G$, which determines the optimal perturbations under the transient growth formalism, at time $t=\tau$ is given by
\begin{equation}
\begin{aligned}
    G(\tau) & \equiv \sup_{\Tilde{\bm{u}}(0) \neq \bm{0}} \frac{E(\Tilde{\bm{u}}(\tau))}{E(\Tilde{\bm{u}}(0))} = \sup_{\Tilde{\bm{\xi}}_0 \neq \bm{0}} \frac{ \big\lVert \mathsfbi{F} \exp (\tau \mathsfbi{S}) \Tilde{\bm{\xi}}_0 \big\rVert_2^2}{ \big\lVert \mathsfbi{F} \Tilde{\bm{\xi}}_0 \big\rVert_2^2}
    \\ & =  \big\lVert \mathsfbi{F} \exp (\tau \mathsfbi{S}) \mathsfbi{F}^{-1} \big\rVert_2^2. &
    \label{max-energy-growth}
\end{aligned}
\end{equation}
Using the fact that the $L^2$ norm of an arbitrary matrix is the same as its largest singular value, we finally reach the following: For the largest singular value $\varsigma_{1}$ (assumed to be non-zero) of $\mathsfbi{F} \exp (\tau \mathsfbi{S}) \mathsfbi{F}^{-1}$ and its associated right and left singular vectors $\bm{r}_{1}$ and $\bm{l}_{1}$, i.e.,
\begin{equation}
    \mathsfbi{F} \exp (\tau \mathsfbi{S}) \mathsfbi{F}^{-1} \bm{r}_{1} = \varsigma_{1}\bm{l}_{1}, 
    \label{svd}
\end{equation}
which can result from the singular value decomposition (SVD), we get
\begin{equation}
    G(\tau) = \varsigma_{1}^2,
    \label{max-energy-growth-svd}
\end{equation}
and the optimal perturbation velocity input and output at time $t = \tau$ are
\begin{equation}
    \Tilde{\bm{\upsilon}}_{\text{opt}}(0) = \mathsfbi{P} \mathsfbi{V} \mathsfbi{F}^{-1} \bm{r}_{1}
    ~~~\text{and}~~~ 
    \Tilde{\bm{\upsilon}}_{\text{opt}}(\tau) = \varsigma_{1}\mathsfbi{P} \mathsfbi{V} \mathsfbi{F}^{-1} \bm{l}_{1} .
    \label{opt-perturbations}
\end{equation}

\subsection{Numerical parameters}\label{sec:numericalparameters}
Discretisation is essential for the current problem formulation, and care must be taken to minimise non-physical errors arising from numerical parameters. In this study, we employ the mapped Legendre spectral collocation method, which is well-suited for analysing rotating flows in unbounded domains. The key numerical parameters in this scheme include the number of spectral basis elements $M$, the number of radial collocation points $N$, and the map parameter $L$ — defined in \eqref{mapparam}. Detailed procedures for accurately resolving the eigenmodes are discussed in \citet{Lee2023}, which interested readers are encouraged to read. In our setup, we keep $M + |m| = N$ to ensure $N \geq M$, and we vary $L$ to explore both potential and viscous critical-layer eigenmode families by adjusting the scheme's characteristic numerical resolution.

For comparison, we secondarily consider the Chebyshev spectral collocation method with domain truncation at $r = R_\infty$. The domain of the Chebyshev polynomials is linearly mapped from $[-1, 1]$ to $[0, R_\infty]$, as favoured in previous studies \citep[e.g.,][]{Khorrami1989, Mao2011}, where the primitive variables $\Tilde{\bm{u}}$ and $\Tilde{p}$ are taken as state variables. However, the use of this scheme in the present study is solely restricted to investigating how sensitively the domain truncation affects the transient growth analysis outcome, in comparison to the mapped Legendre spectral collocation method developed for radially unbounded domains \citep{Matsushima1997, Lee2023}, serving as a significant drawback despite its constructional convenience for computation.

\subsection{Physical parameters}\label{sec:physicalparameters}
There are five physical parameters that influence the nature of the problem: $\tau$, $q$, $\Rey$, $m$ and $\kappa$. Below, we clarify the range or value of each parameter that will be the focus of this study.

The maximum energy growth $G$ is explicitly dependent on the total time of growth $\tau$, indicating the duration over which we allow linear transient dynamics of the wake vortex to develop. In the context of aircraft trailing vortices, an upper limit on $\tau$ is identifiable due to the dominance of the Crow instability mechanism after several hundred time units ($= R_0/U_0$). For example, \citet[][pp. 341-343]{Matsushima1997} reported the prevalence of long-wavelength instability around $t = 200$ in simulations of a counter-rotating vortex pair configuration. Under proper rescaling of units, the trailing vortex simulation by \citet[][pp. 295-297, also see figure 10]{Han2000} exhibited vortex linkage at $t = 229.12$ under moderate ambient turbulence. Based on these findings, we concentrate on the region where $\tau \lesssim O(10^2)$. Our typical attention to the transient growth is in the time range of $10 < \tau \le 100$, where relatively fast transient growth is expected \citep{Mao2012}. However, we note that a longer range may be explored in case it is needed to verify long-term characteristics of transient growth.

Additionally, the analysis outcomes are also subject to physical parameters such as the swirl parameter $q$, the Reynolds number $\Rey$ and the azimuthal and axial wavenumbers $m$ and $\kappa$. For this study, we fix the first two parameters and, unless specified otherwise, use $q = 4$ and $\Rey = 10^5$. These values represent conditions where the swirling motion is sufficiently strong to exclude significant linear instabilities \citep[e.g., $q \ge 2.31$, see][]{Heaton2007Centre}, and where viscous diffusion is small enough to treat the base vortex profile as quasi-steady. It is remarked that, according to experiment-based estimation by \citet[][p. 259]{Fabre2004}, this setup aligns with the condition of actual trailing vortices behind large transport aircraft. As for the perturbation wavenumbers, we take attention to axisymmetric or helical cases ($m=0~\text{or}~1$) with small axial wavenumbers of order unity or less. This choice is driven not only by their prevalance in vortex transient growth literature \citep[e.g.,][]{Antkowiak2004, Pradeep2006, Mao2012, Navrose2018}, but also by the anticipation that such low-frequency perturbations better account for the principal perturbation structure that we later aim to initiate via particles near the vortex.

\section{Optimal perturbations}\label{sec:optimalperturbations}
\subsection{Numerical sensitivity and proper discretisation}\label{sec:numericalsensitivity}
We construct optimal perturbations by combining the eigenmodes of the wake vortex. To obtain accurate results, the chosen computation scheme should reliably capture each physically relevant eigenmode family in a well-resolved manner, while maintaining insensitivity to variations in numerical parameters. We evaluate the numerical sensitivity of the mapped Legendre spectral collocation method and the Chebyshev spectral collocation method to determine which is more appropriate for the present analysis. 

When considering viscous eigenmodes that are regular across the entire radial domain, including their asymptotic behaviours near the origin and as $r$ approaches infinity, there are three important eigenmode families: the discrete family, the potential family, and the viscous critical-layer family \citep{Lee2023}. The first family, as its name suggests, is associated with discrete spectra (i.e., sets of eigenvalues), and each eigenmode's spatial structure is uniquely characterised by the number of `wiggles' clustered in or around the vortex core. The other two families comprise continuous eigenmodes, whose spatial structures vary continuously and are associated with continuous spectra.

\begin{figure}
    \vspace{.1in}
    \centerline{\includegraphics[width=\textwidth,keepaspectratio]{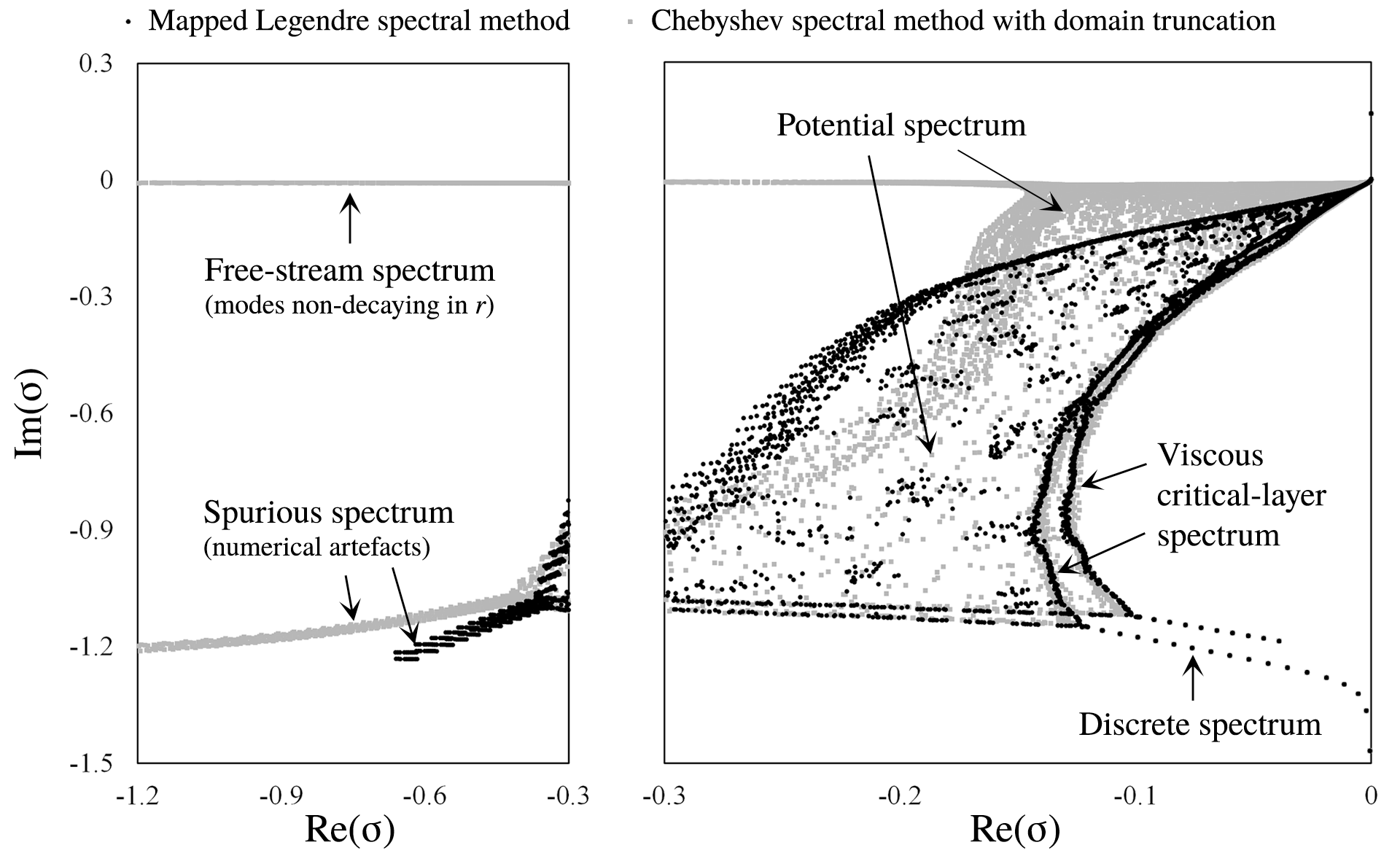}}
    \caption{Numerical spectra of the $q$-vortex with $(m,\,\kappa,\,q,\,\Rey) = (1,\,1.0,\,4.0,\,10^5)$ using the Chebyshev spectral collocation method (grey squares) and the mapped Legendre spectral collocation method (black dots). The consistent discrete spectrum demonstrates the robustness of both methods. The free-stream and spurious spectra (shown in the left panel) are associated with either singular or non-physical eigenmodes, making them irrelevant to this study, with most generated by the Chebyshev method. The discrete, potential and viscous critical-layer spectra (shown in the right panel) are associated with regular eigenmodes, with the mapped Legendre method providing a clearer distinction of the viscous critical-layer spectrum.}
\label{fig:cheb_vs_leg}
\end{figure}

Figure \ref{fig:cheb_vs_leg} shows the numerically resolved spectra of the $q$-vortex using the following physical parameters: $(m,\,\kappa,\,q,\,\Rey) = (1,\,1.0,\,4.0,\,10^5)$, envisioning the families of eigenmodes. Both the Chebyshev spectral collocation method (with  $M=800$) and the mapped Legendre spectral collocation method (with $M=400$) were employed for this analysis. To illustrate the continuous spectra, we collected all numerical eigenvalues obtained by varying the domain truncation radius $R_\infty$, ranging from $12$ to $13$ for the Chebyshev spectral method, and by adjusting the map parameter $L$, spanning from $3$ to $3.1$ for the mapped Legendre spectral method. These variations in $R_\infty$ or $L$ introduce small shifts in the numerically resolved continuous eigenvalues, making it possible to trace continuous spectra \citep[see][\S 6.4.2]{Lee2023}.

Given perfect resolution, the free-stream and potential spectra are expected to stretch out to $\Real(\sigma) \rightarrow -\infty$. In figure \ref{fig:cheb_vs_leg}, the spectra are shown in two panels with different aspect ratios. The left panel extends to large $|\Real(\sigma)|$, showcasing both the free-stream and spurious spectra. The eigenmodes related to the free-stream spectrum and the spurious spectrum exhibit non-regular characteristics: the free-stream eigenmodes are singular since they do not decay to zero as $r \rightarrow \infty$ \citep{Mao2011}, and the spurious eigenmodes are non-physical, characterised by irregular oscillations near the origin \citep{Lee2023}. On the other hand, the right panel displays the discrete, potential, and viscous critical-layer spectra, which correspond to the regular eigenmode families discussed earlier.

A few issues arise when using the Chebyshev spectral method instead of the mapped Legendre spectral method for resolving the eigenmodes. As depicted in the left panel of figure \ref{fig:cheb_vs_leg}, a significant portion of the numerically resolved spectra accounts for eigenmode families that are either singular or non-physical, making them irrelevant to the present problem. This issue likely arises from approximating the asymptotic constraints through subordinate boundary conditions at both ends of the computational domain. For the Chebyshev spectral method, as for $m = 1$, the boundary conditions implemented are
\begin{equation}
    \frac{d \Tilde{u}_r}{dr} = \frac{d \Tilde{u}_\phi}{dr} = \Tilde{u}_z = \Tilde{p} = 0~~~~\text{at}~~r=0,
    \label{originbc}
\end{equation}
and
\begin{equation}
    \Tilde{u}_r = \Tilde{u}_\phi = \Tilde{u}_z = \Tilde{p} = 0~~~~\text{at}~~r=R_\infty,
    \label{inftybc}
\end{equation}
which are proxies for analyticity at the origin and rapid decay as $r \rightarrow \infty$, respectively \citep[see][]{Ash1995}. While \eqref{originbc} and \eqref{inftybc} may serve as necessary conditions for what they are supposed to mimic, they cannot be considered formally equivalent. For instance, \eqref{inftybc} does not prohibit solutions from oscillating in the far field as long as the oscillation is momentarily zeroed out at $r=R_\infty$, which explains the emergence of the free-stream spectrum.

\begin{figure}
    \vspace{.1in}
    \centerline{\includegraphics[width=\textwidth,keepaspectratio]{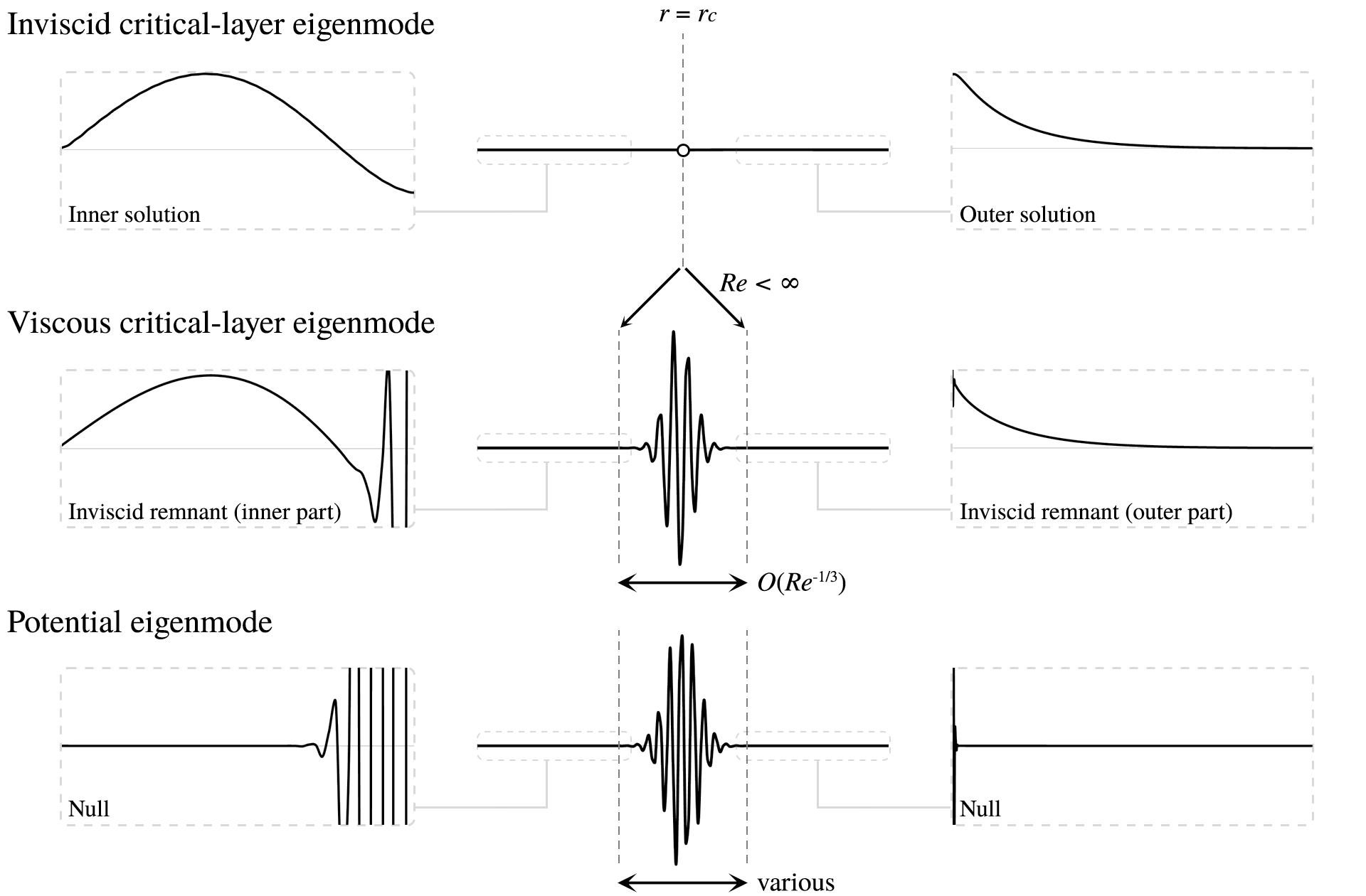}}
    \caption{Schematic comparison between viscous critical-layer and potential eigenmodes, illustrating a velocity component. Both exhibit a similar large-scale structure within the region where viscosity effects are locally dominant, corresponding to a singularity at $r=r_c$ in the inviscid limit \citep[for details of this inviscid singularity, see][]{Lee2023}. The width of this large-scale structure scales as $O(Re^{-1/3})$ \citep[see][]{Lin1955} for the viscous critical-layer eigenmode. In contrast, the width of the potential eigenmode can vary even at the same $Re$, forming a `wave packet' in compliance with the pseudomode analysis \citep[see][]{Trefethen2005}. Aside from the location $r=r_c$, the viscous critical-layer eigenmode retains the structure of its inviscid counterpart, while the potential eigenmode simply turns into null.}
\label{fig:pot_vs_visc}
\end{figure}

The second issue comes from the unclear distinction between the viscous critical-layer spectrum and the potential spectrum. As noted in \citet[][pp. 41-42]{Lee2023}, the Chebyshev spectral method produces scattered traces of the viscous critical-layer spectrum curves, making it challenging to distinguish these curves from the surrounding continuous region. This scattering can be attributed to the high sensitivity of continuous spectra to minor errors. In the Chebyshev spectral method, domain truncation removes spatial information far from the origin, which, albeit diminutive, holds physical significance. Figure \ref{fig:pot_vs_visc} illustrates the difference between the viscous critical-layer eigenmodes and the potential ones. Despite their structural resemblance on a large scale, the viscous critical-layer eigenmodes retain the structure of their inviscid counterparts beyond the region where viscosity effects dominate locally, scaled in the order of $Re^{-1/3}$ \citep[][]{Lin1955}. In contrast, the potential eigenmodes turns into null outside this region, epitomising their `wave packet' form \citep{Mao2011}, which conforms to the twist condition presented by \citet[][pp. 98-114]{Trefethen2005}. Further details of their comparison are omitted in this article; they are elucidated in \citet{Lee2023}.

\begin{figure}
    \vspace{.1in}
    \centerline{\includegraphics[width=\textwidth,keepaspectratio]{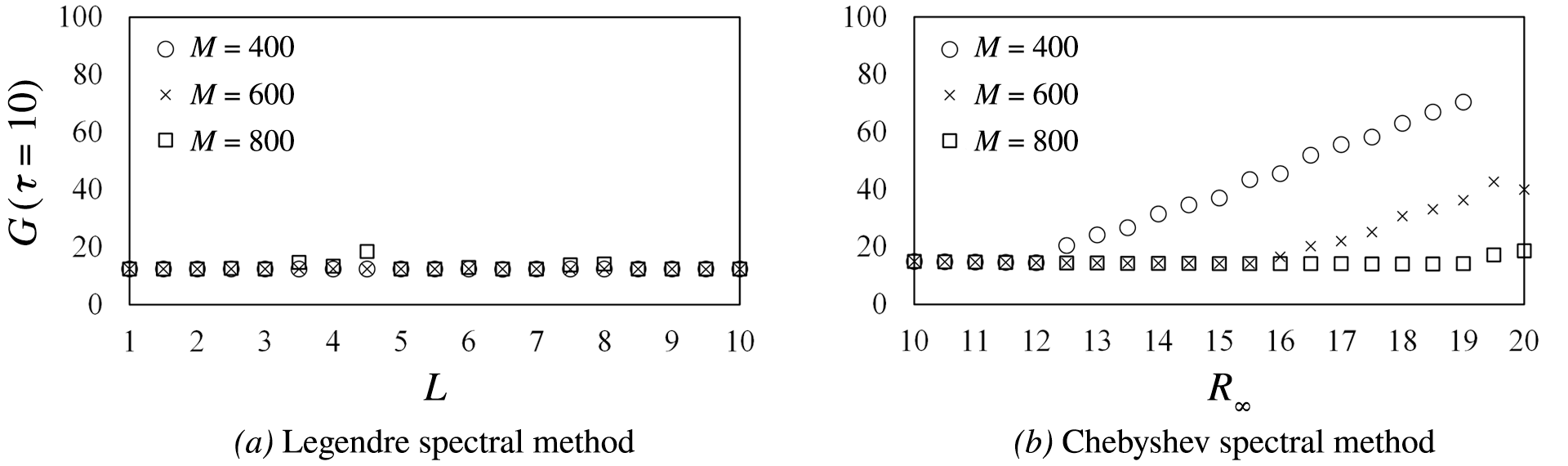}}
    \caption{Numerical sensitivity test evaluating $G(\tau = 10)$ for the entire eigenspace associated with $(m,\,\kappa,\,q,\,\Rey) = (1,\,1.0,\,4.0,\,10^5)$ using $(a)$ the mapped Legendre spectral collocation method with varying $L$, and $(b)$ the Chebyshev spectral collocation method with varying $R_\infty$, for $M=400,\,600$ and $800$. The test ranges reflect typical parameter usage. $L$ can be arbitrarily adjusted without impacting domain unboundedness. In contrast, setting a finite $R_\infty$ compromises the unbounded nature of the problem, and therefore, larger values of $R_\infty$ should be preferred. However, increasing $R_\infty$ leads to undesirable numerical sensitivity.}
\label{fig:sensitivity}
\end{figure}

A numerical sensitivity test evaluating the maximum energy growth $G$ at $\tau = 10$ for the entire eigenspace, using both methods, is presented in figure \ref{fig:sensitivity}. The remaining physical parameters are kept the same: $(m,\,\kappa,\,q,\,\Rey) = (1,\,1.0,\,4.0,\,10^5)$. As expected, increasing the number of spectral elements $M$ reduces sensitivity to changes in numerical parameters for both methods. At a fixed $M$, the map parameter $L$ acts as a resolution tuning parameter in the mapped Legendre spectral collocation method, while the domain truncation radius $R_\infty$ serves this role in the Chebyshev spectral collocation method. The parameter test ranges shown in figure \ref{fig:sensitivity} are based on the typical usage found in \citet{Lee2023} and \citet{Mao2011}. Changes in $L$ have minimal impact on $G(\tau = 10)$ within the mapped Legendre spectral collocation method's test range ($1 \le L \le 10$), therefore allowing for arbitrarily selection of $L$ within this interval. In contrast, $G(\tau = 10)$ is notably influenced by variations in $R_\infty$ within the Chebyshev spectral collocation method's test range ($10 \le R_\infty \le 20$), especially as $R_\infty$ increases. This presents challenge, as using a large $R_\infty$ should be preferred to preserve the unbounded nature of the radial domain. We found that manually excluding the sub-eigenspace spanned by free-stream eigenmodes mitigates this issue, which is, in fact, a step that is proactively taken in the mapped Legendre spectral collocation method. Exclusion of these eigenmodes is reasonable from a physical standpoint, as their non-decaying behaviour implies that they analytically possess infinite energy. This, in turn, renders them formally inapplicable in the current transient growth analysis context.

To recapitulate, when it comes to resolving the eigenmodes of the $q$-vortex in a radially unbounded domain, the Chebyshev spectral collocation method with domain truncation faces several challenges, which unfavourably influence numerical sensitivity in transient growth evaluation. In contrast, the mapped Legendre spectral collocation method effectively mitigates these numerical limitations, making it a more suitable choice for the present problem. Therefore, we adopt the mapped Legendre spectral collocation method for examining the transient growth of the wake vortex.

\subsection{Maximum energy growth}\label{sec:maximumenergygrowth}
\citet{Mao2012} demonstrated that transient growth primarily results from the non-normality of continuous eigenmodes, while discrete eigenmodes play a less significant role. In their analysis, they used the term `continuous eigenmodes' as a compilation of potential and free-stream eigenmodes. However, as we pointed out earlier, free-stream eigenmodes are unsuitable for evaluating maximum energy growth because their energy reaches infinity. Thus, the term `continuous eigenmodes' in their argument should more specifically refer to potential eigenmodes. Not only that, but their argument also requires further refinement, as it did not account for viscous critical-layer eigenmodes. This eigenmode family was not distinguished from the potential family, presumably due to spectral overlap between the two \citep[see][p. 8]{Mao2011} and their large-scale structural similarity, as shown in figure \ref{fig:pot_vs_visc}. 

Accordingly, we believe that the argument put forth by \citet{Mao2012} still necessitates clarification about which continuous eigenmode family predominantly contributes to optimal perturbations that maximise energy growth: the potential family or the viscous critical-layer family. To that end, we first evaluate $G (\tau)$ across the entire eigenspace and then compare the results with those from different sub-eigenspaces, each spanned by a distinct eigenmode family.

\begin{figure}
    \vspace{.1in}
    \centerline{\includegraphics[width=\textwidth,keepaspectratio]{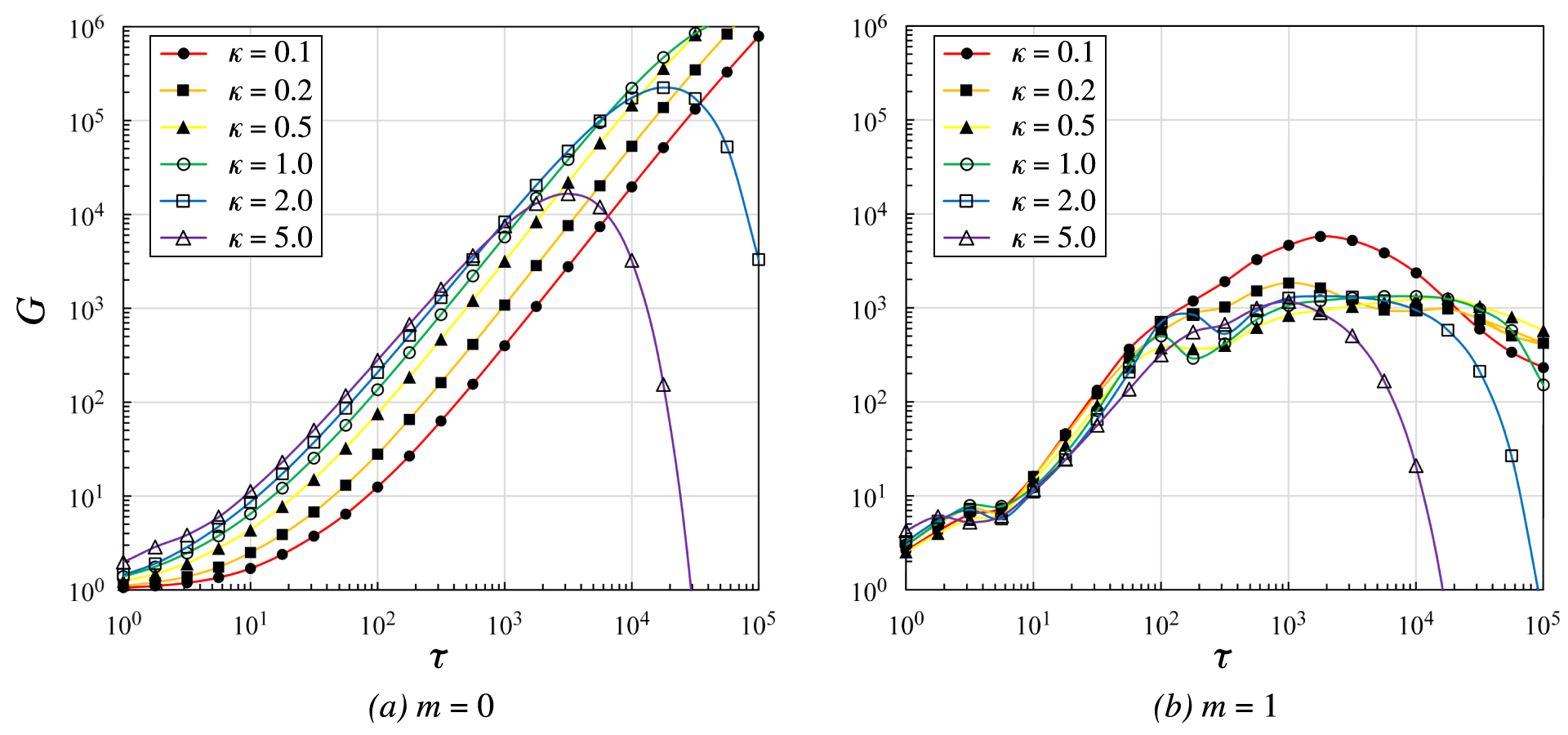}}
    \caption{Maximum energy growth as a function of total growth time: $(a)$ for $m=0$ (axisymmetric) and $(b)$ for $m=1$ (helical). The values of $G$ are evaluated from the entire eigenspace, incorporating the discrete, potential and viscous critical-layer families as basis elements. Here, $q=4$ and $\Rey = 10^5$.}
\label{fig:transient_growth_g}
\end{figure}

Figure \ref{fig:transient_growth_g} presents the numerically evaluated values of $G(\tau)$ from the entire eigenspace at various wavenumbers $\kappa$ for the $m=0$ and $m=1$ cases. In the $m=0$ cases, as shown in figure \ref{fig:transient_growth_g}$(a)$, where perturbations are two-dimensional (i.e., functions of $r$ and $z$ only), the dependence of the $G$ curves on $\kappa$ is clear; in the short run, growth is stronger with larger $\kappa$, while in the long run, the largest $G$ is achieved with smaller $\kappa$. One may check in figure \ref{fig:transient_growth_g}$(a)$ that the upper envelope of the curves sequentially corresponds to decreasing $\kappa$ as $\tau$ increases. This trend aligns with previous observations in the literature \citep{Pradeep2006, Mao2012}, supporting the validity of our evaluation.

\begin{figure}
    \vspace{.1in}
    \centerline{\includegraphics[width=5.2in,keepaspectratio]{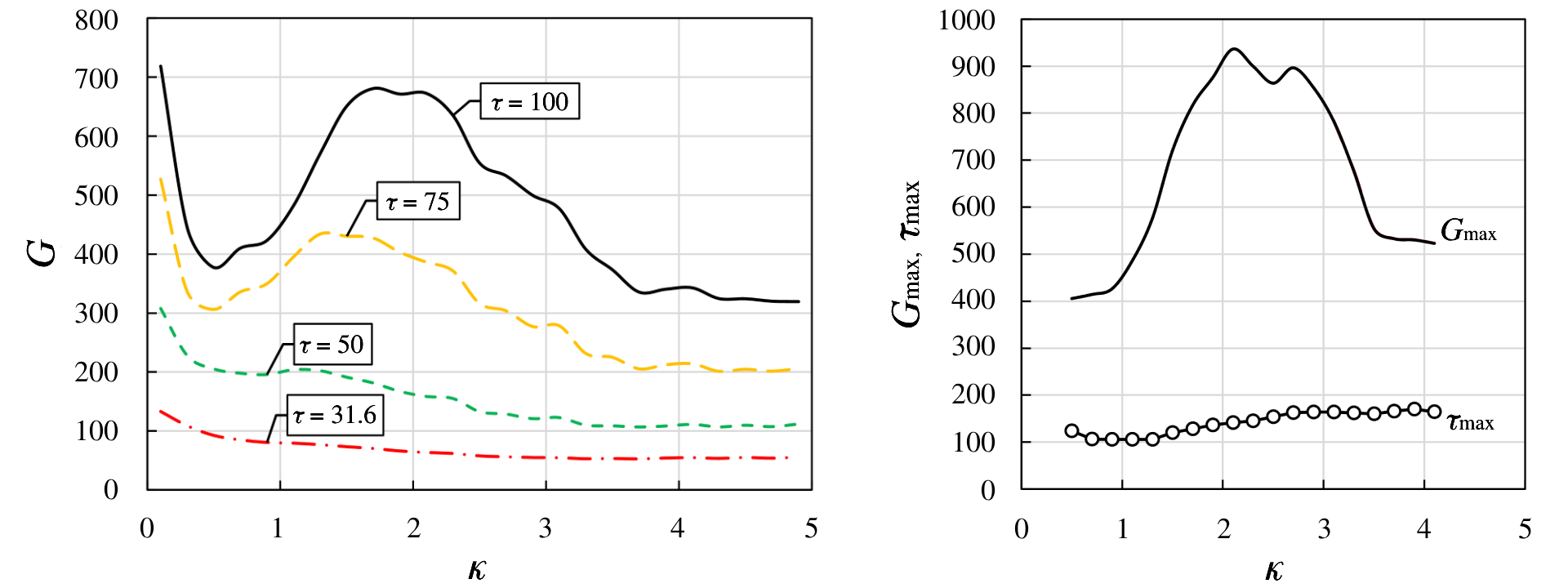}}
    \caption{(Left) maximum energy growth as a function of axial wavenumber for $m=1$, at specified growth periods of $\tau = 31.6,~50,~75$ and $100$, and (right) local maximum of $G$ across growth time around $\tau = 100$ and corresponding growth time. As in figure \ref{fig:transient_growth_g}, the values of $G$ are evaluated from the entire eigenspace. Here, $q=4$ and $\Rey = 10^5$.}
\label{fig:vortevoldifftaus}
\end{figure}

In the $m=1$ cases, depicted in figure \ref{fig:transient_growth_g}$(b)$, involving three-dimensional perturbations, the $\kappa$-dependence of the $G$ curves becomes complex, as previously noted by \citet{Antkowiak2004} and \citet{Pradeep2006}. To further clarify this trend, figure \ref{fig:vortevoldifftaus} (left panel) presents supplementary slices of $G$ as a function of axial wavenumber $\kappa$ for $m=1$ at four growth periods within the time range of interest: $\tau=31.6$, $50$, $75$ and $100$. As $\tau$ increases, a local energy growth peak becomes more pronounced, especially at $\tau=100$, where the local peak around $\kappa = 2.0$ nearly stands at the largest $G$ at $\kappa = 0.1$. This peak feature in the $G$-$\kappa$ curves aligns with the findings of \citet[][p. L3]{Antkowiak2004}, who ascribed the intricate (stretching and tilting) nature of three-dimensional perturbations to such irregularities.

In the right panel of figure \ref{fig:vortevoldifftaus}, the local maximum of $G$ around $\tau = 100$, denoted $G_{\mathrm{max}}$, is plotted alongside its growth time, $\tau_{\mathrm{max}}$, as a function of $\kappa$, provided this maximum is identifiable (e.g., see the $\kappa = 1.0$ or $2.0$ cases in figure \ref{fig:transient_growth_g}$(b)$). This approach follows \citet{Antkowiak2004, Pradeep2006}, who used $G_{\mathrm{max}}$ to examine local energy growth features. Similar to the $G-\kappa$ curve at $\tau=100$, the local energy growth peak is observed , as $\tau_{\mathrm{max}}$ consistently appears near $\tau = 100$. Note that we avoid using the `global' maximum of $G$ over the entire range of $\tau$, as suggested in the literature. Due to the higher order of magnitude of $\Rey(=10^5)$ in the present study compared to that in the literature ($\lesssim 10^4$), the global maximum of $G$ occurs at a much larger growth time, $\tau \sim O(10^3-10^4)$. This time range far exceeds the intervals considered both in the literature and our study ($\tau \lesssim 10^2$), thus placing it outside the current scope of analysis. 

When comparing the $m=1$ cases to the $m=0$ cases, focusing on relatively short-term growth, we find that the largest $G$ for $m=1$ generally exceeds that for $m=0$. For example, at $\tau = 10^2$, the largest $G$ among the evaluated values is $7.2 \times 10^2$ for $\kappa = 0.1$ at $m=1$, whereas it is $2.8 \times 10^2$ for $\kappa = 5.0$ at $m=0$.

\begin{figure}
    \vspace{.1in}
    \centerline{\includegraphics[width=\textwidth,keepaspectratio]{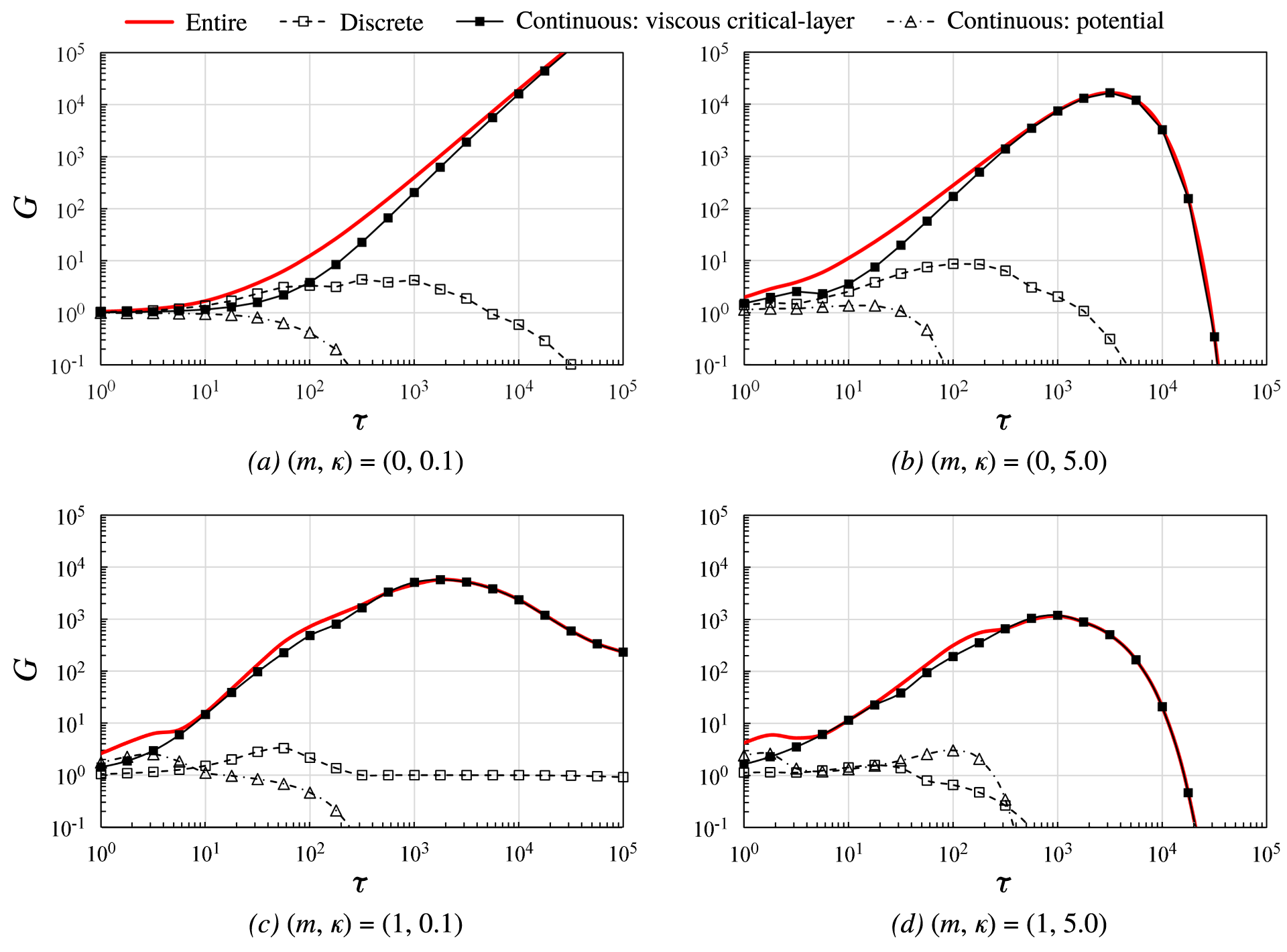}}
    \caption{Comparison of the curves of $G(\tau)$ evaluated from different sub-eigenspaces, each spanned by a distinct eigenmode family: $(a)$ $(m,\,\kappa) = (0,\,0.1)$, $(b)$ $(m,\,\kappa) = (0,\,5.0)$, $(c)$ $(m,\,\kappa) = (1,\,0.1)$, and $(d)$ $(m,\,\kappa) = (1,\,5.0)$. Here, $q=4$ and $\Rey = 10^5$. The maximum energy growth curves from the entire eigenspace, identical to those plotted in figure \ref{fig:transient_growth_g}, are nearly reproduced by those from the sub-eigenspace spanned by the viscous critical-layer family.}
\label{fig:transient_growth_subg}
\end{figure}

The maximum energy growth curves, evaluated from the entire eigenspace and compared with those from sub-eigenspaces respectively spanned by the discrete family, the viscous critical-layer family and the potential family, are presented in figure \ref{fig:transient_growth_subg}. It is evident that the curves derived from the entire eigenspace are primarily reproduced by those obtained from the sub-eigenspace of the viscous critical-layer family, highlighting its dominant contribution. On the other hand, the values of $G(\tau)$ from the rest of the continuous sub-eigenspace, for which the potential eigenmodes account, are of a similar magnitude to those from the discrete sub-eigenspace. Thus, the contribution of the potential family to transient growth is as minor as that of the discrete family. 

There are two minor exceptions worth noting. For $m=0$, the discrete family contributes as significantly as the viscous critical-layer family to short-term optimal growth, particularly when $\kappa$ is small. We believe that this is relevant to the fact that, in the limit of $\kappa \rightarrow 0$, critical layers vanish and so do the derived continuous eigenmodes, while the discrete ones persist. Additionally, for $m=1$, the potential family's contribution to optimal growth slightly supersedes that of the viscous critical-layer family during a brief period ($\tau < 10$), accounting for the presence of a quirk in the $G$ curves around $\tau = 5.6$. However, the exception clears quickly beyond this period, and the maximum $G$ attainable during this period never exceeds $10$, thus not overturning the general dominance of the viscous critical-layer family in transient growth.

Based on these observations, we revisit the demonstration provided by \citet{Mao2012} with the following clarification; the non-normality of the continuous eigenmodes induces significant transient growth, and it is specifically the viscous critical-layer family that predominantly contributes to this growth, rather than the potential family. The distinction is important, as it addresses the `true' origin of transient growth of the wake vortex as critical layers. The potential eigenmodes have their theoretical root in the wave packet pseudomode analysis \citep{Trefethen2005}. As showcased in figure \ref{fig:pot_vs_visc}, they omit the asymptotic information of critical layers, making their birth irrelevant to phenomena that require asymptotic matching or equivalently, critical layer analysis \citep{Lin1955, LeDizes2004}. Although the wave packet pseudomode analysis is a powerful tool for exploring all possible forms of continually varying eigensolutions, it can divert attention too much from the genuine gems more worthy of our focus. Furthermore, this clarification better aligns with the argument made by \cite{Heaton2007Optimal}, who suggested that inviscid continuous spectrum (CS) transients dominate growth over short time intervals. The viscous critical-layer family in our classfication corresponds to the viscous regularisation of the inviscid CS, which we denoted as the inviscid critical-layer spectrum \citep{Lee2023}, when $\Rey < \infty$.

\subsection{Perturbation structures}\label{sec:perturbationstructures}
The effects of chaging $m$ and $\kappa$ on the perturbation structures that lead to optimal transient growth have been widely investigated and are well-established in the context of linear vortex dynamics \citep{Antkowiak2004, Pradeep2006, Mao2012}. In this section, we examine whether our transient growth calculation complies with these established findings and then conduct a comparative analysis of the perturbation structures across different values of $m$ and $\kappa$ for further consideration.

\citet{Pradeep2006} reported that axisymmetric perturbations ($m=0$) generally produce the largest energy growth, as illustrated in figure \ref{fig:transient_growth_g}. However, as the largest $G$ increases, the total duration of perturbation growth also lengthens (i.e., the $\tau$ needed to achieve $G$ increases), as the spatial structure shifts further from the vortex core, necessitating longer time for interactions to occur. For helical perturbations ($m=1$), a common spatial structure emerges, with the main motion concentrated around a specific radius near the vortex core. In these cases, the growth can potentially trigger fluctuations within the vortex core, even though the initial perturbation originates outside it. This mechanism has occasionally been identified as a cause of erratic long-wavelength displacements in experimental vortices \citep[e.g.,][]{Edstrand2016, Bolle2023}, often termed `vortex meandering' \citep[see][p. L4]{Antkowiak2004}. \citet{Mao2012} affirmed that the vortex meandering phenomenon can be driven by the transient response of the vortex to an out-of-core perturbation.

\begin{figure}
    \vspace{.1in}
    \centerline{\includegraphics[width=\textwidth,keepaspectratio]{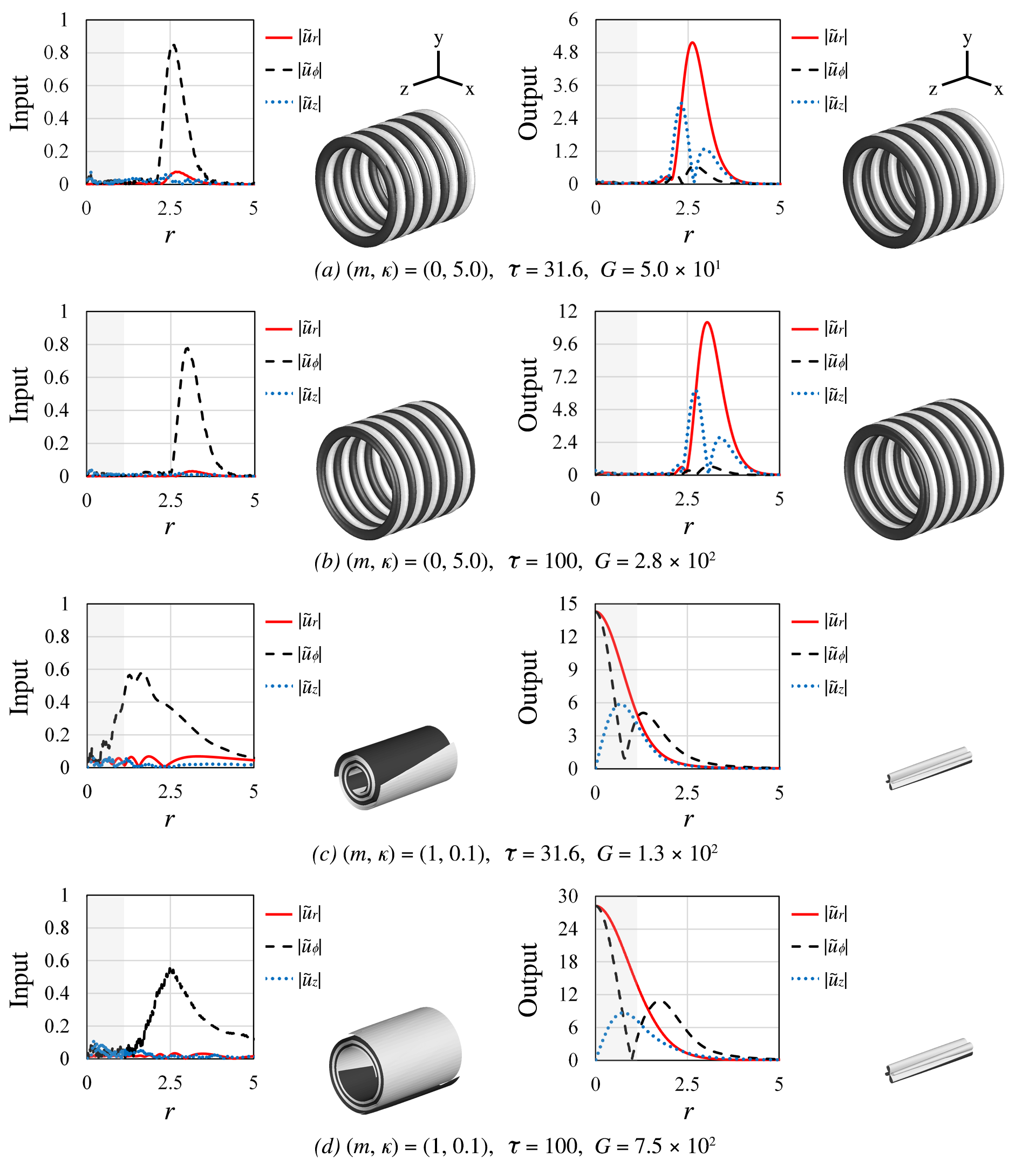}}
    \caption{Optimal perturbation inputs with unit energy $E$ (as defined in \eqref{energy-integ}) and their amplified outputs at $t=\tau$, shown through absolute velocity components alongside their corresponding three-dimensional structures. Each structure is represented by the isosurface at 50\% of the maximum specific energy in space. Dark and light colours indicate counterclockwise and clockwise swirling of the flow, respectively. Here, four representative cases with the largest value of $G$ from figure \ref{fig:transient_growth_g} are displayed: $(a,\,b)$ the axisymmetric cases $(m,\,\kappa) = (0,\,5.0)$ for $\tau = 31.6$ and $\tau = 100$, respectively, and $(c,\,d)$ the helical cases $(m,\,\kappa) = (1,\,1.0)$ for $\tau = 31.6$ and $\tau = 100$, respectively. The initial dominance of azimuthal velocity components is found in all cases, while for $m=1$, energy is transferred into the core region $r \le 1.12$ (shaded in each plot).}
\label{fig:pert_struct}
\end{figure}

As mentioned in \S \ref{sec:physicalparameters}, our focus is on the relatively short time period of $10 < \tau \leq 100$ to study the transient growth process. In longer periods, classical linear instability mechanisms like the Crow instability may dominate under real conditions. Within this time range, the largest values of $G$ attained from our considerations (see figure \ref{fig:transient_growth_g}) occur at $\kappa = 5.0$ for the $m=0$ cases and at $\kappa = 0.1$ for the $m=1$ cases. We consider these cases as representative. In figure \ref{fig:pert_struct}, we depict the optimal perturbation velocity inputs and outputs for $\tau = 31.6$ and $\tau=100$, all of which are visualised by the absolute velocity components. For clearer visualisation, we portray their corresponding three-dimensional structures alongside, represented by the iso-surface at 50\% of the maximum specific energy in physical space, i.e., $|(\Tilde{\bm{u}}(r) e^{\mathrm{i}(m \phi + \kappa z)} + \mathrm{c.c.})|^2/2$, where $\mathrm{c.c.}$ stands for the complex conjugate of the antecedent term. Dark and light surfaces express counterclockwise and clockwise swirling directions, respectively.

In all cases, the following characteristics are consistently observed. First, azimuthal velocity components are initially dominant in all optimal perturbations, while the other velocity components evolve significantly towards the end of the growth period. This clearly indicates that the azimuthal velocity component should be prioritised when inducing these optimal perturbations from an unperturbed state. Second, the most energetic part of the optimal perturbation inputs, coinciding with the peak of the absolute azimuthal velocity component, tends to be distant from the vortex core as $\tau$ increases. This tendency is found to be more evident in cases $(c)$ and $(d)$, where the major perturbation structure overlaps the core region at $\tau = 31.6$ but moves out of the core at $\tau = 100$.

For cases $(a)$ and $(b)$, where $(m,\,\kappa)=(0,\,5.0)$, the input perturbations generally form a ring structure owing to their azimuthal symmetry. As the perturbation evolves, the radius of the ring remain largely unchanged. This tendency for local confinement of the optimal perturbation structures becomes more pronounced with increasing $\kappa$. This indicates that perturbations with shorter axial wavelengths have a more localised influence around the initially perturbed region.

For cases $(c)$ and $(d)$, where $(m,\,\kappa)=(1,\,0.1)$, a spiral structure develops in the most energetic region of the input perturbation due to alternating layers of oppositely swirling fluid motions at the periphery of the vortex core. Unlike the axisymmetric cases, the perturbation structure undergoes a drastic transformation from its input to output states. Notably, the most energetic region of the perturbation, initially located outside the vortex core, eventually penetrates into the vortex core. During this process, the transverse velocity ($\Tilde{u}_r$ and $\Tilde{u}_\phi$) becomes maximal at the vortex centre. In other words, the principal response of the vortex to optimal perturbations with $m=1$ is characterised by the transverse motion of the vortex core, which is likely linked to the vortex meandering phenomenon \citep{Edstrand2016, Bolle2021}.

It is important to clarify that the induction of vortex meandering by optimal helical perturbations with an axially long wavelength was formerly given by \citet{Mao2012}. They employed mesh-based direct numerical simulations rather than the matrix-based analysis (corresponding to \eqref{ode-disc} - \eqref{opt-perturbations} in our formulation), even though they used a matrix-based approach when $m=0$. In a way, this choice seems to have been made to address challenges related to analyticity at the origin, which depends on the value of $m$ \citep[see][pp. 51-52]{Lee2023}. In contrast, our approach, utilising the mapped Legendre spectral collocation method, is fundamentally designed to be robust for any value of $m$. Therefore, our contribution here lies in confirming the same phenomenon linked to $m=1$ perturbations using a computationally fast and formally consistent matrix-based transient growth analysis.

\subsection{Non-linear impacts on an optimally perturbed vortex}\label{sec:nonlinearimpacts}
Given an optimally perturbed vortex, the linearised theory (see \S \ref{sec:formulation}) predicts that the perturbation will gradually amplify as time approaches $t=\tau$, after which it decays in the absence of extrinsic factors capable of triggering secondary instabilities from the most perturbed state. This section is dedicated to verifying whether such transient behaviour remains significant in the original non-linear system, governed by \eqref{nonlingoveqn}, despite the influence of higher-order energy transfer across different wavenumbers and other non-linear effects, which could potentially cause early vortex growth to deviate from the linear prediction.

\begin{figure}
    \vspace{.1in}
    \centerline{\includegraphics[width=2.in,keepaspectratio]{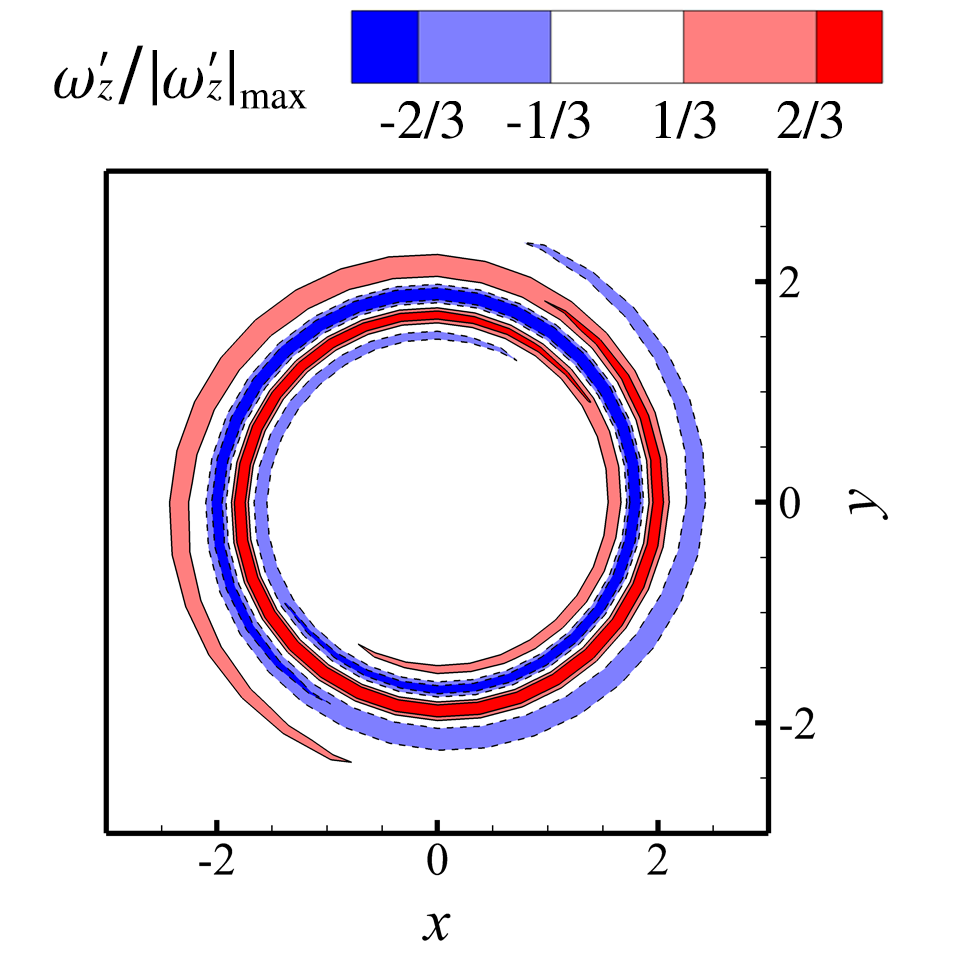}}
    \caption{Axial perturbation vorticity contour on the $z=0$ plane of the optimal input for the case where $(m,\,\kappa) = (1,\,0.1)$ with the optimal growth time $\tau=50$. Solid contour lines represent positive levels (67 \% and 33 \% of the absolute maximum) and dashed lines indicate negative levels (-33 \% and -67 \% of the absolute maximum). }
\label{fig:vortinputz=0}
\end{figure}

Although the optimal perturbation structures vary with the selection of $m$, $\kappa$, and $\tau$, our primary aim here is to investigate their general trend of evolution over time, as anticipated by the linearised theory. We focus on a specific case of optimal perturbation where $(m\,,\kappa) = (1\,,0.1)$ and the optimal growth time is $\tau=50$. The $z$-component of the perturbation vorticity input, $\omega'_z (t=0)$, on the $z=0$ plane is illustrated in figure \ref{fig:vortinputz=0}. This case was chosen because the substantial shift of the most energetic portion of the perturbation from the periphery to the vortex core, as shown in figure \ref{fig:pert_struct}$(c,\,d)$, offers a clear illustration of vortex growth. However, we emphasise that this particular behaviour at $m=1$ (potentially related to vortex meandering) is not the primary focus of this study. For readers interested in vortex meandering, we suggest referring to \citet{Edstrand2016} and \citet{Bolle2021}.

According to the linearised theory, the optimal perturbation velocity input, expressed as $\Tilde{\bm{\upsilon}}_{\text{opt}}(0) = \mathsfbi{P} \mathsfbi{V} \mathsfbi{F}^{-1} \bm{r}_{1}$ as in \eqref{opt-perturbations}, evolves at $t=\eta$ as
\begin{equation}
    \Tilde{\bm{\upsilon}}_{\text{opt}}(\eta) = \mathsfbi{P} \mathsfbi{V} \exp (\eta \mathsfbi{S}) \mathsfbi{F}^{-1} \bm{r}_{1}.
    \label{pert_evolve}
\end{equation}
One may check the consistency of the above equation when $\eta = \tau$, with $\Tilde{\bm{\upsilon}}_{\text{opt}}(\tau)$ given in \eqref{opt-perturbations}. We label this prediction from the linearised theory as `linear.' The `linear' prediction, however, may be ideal as it strips off all higher-order interactions coming from the non-linear convection term, i.e., $\bm{u} \times \bm{\omega}$ in \eqref{nonlingoveqn}, which facilitates energy transfer from the perturbation wavenumbers $(m,\,\kappa)$ to their multiples (e.g., $(2m,\,2\kappa)$, $(3m,\,3\kappa),\,\cdots$) or vice versa. We denote the growth of the optimal perturbation, accounting for higher-order interactions, as `non-linear.' The extent of this non-linearity substantially depends on the initial perturbation's energy level. If the perturbation energy approaches zero (or the perturbation is infinitesimal), the `non-linear' evolution should follow the `linear' prediction. We set aside the numerical details of our non-linear simulations in Appendix \ref{appB}.

In the non-linear simulations, the initial velocity field $\bm{u}(t=0)$ is defined as
\begin{equation}
    \bm{u}(r,\phi,z,t=0\,;\,\varepsilon) \equiv \overline{\bm{U}}(r) + \varepsilon \Tilde{\bm{u}}_{\mathrm{opt}}(r) e^{i(m \phi + \kappa z)} + \mathrm{c.c.},
    \label{initcond}
\end{equation}
where $\varepsilon$ determines how intense the initial perturbation is, adjusting the perturbation energy input. The base term representing the unperturbed $q$-vortex, $\overline{U}(r)$, was assumed to be unchanging in time in the linear analysis, as its radial viscous diffusion is negligible due to the high $Re$ number in this problem setup. In contrast, the non-linear simulations take this small viscous diffusion of the base $q$-vortex into account for enhanced accuracy. That is to say, even the unperturbed flow changes slowly over time, which can be calculated with $\varepsilon = 0$. As a result, the perturbation velocity field at $t=\eta$ is assessed as the difference between two time-varying fields, i.e.,
\begin{equation}
    \bm{u}'(r,\phi,z,\eta\,;\,\varepsilon) \equiv \bm{u}(r,\phi,z,\eta\,;\,\varepsilon) - \bm{u}(r,\eta\,;\,0).
    \label{nonlinperteval}
\end{equation}
The perturbation energy at $t=\eta$, denoted $E(\eta)$, is evaluated as the volume integration of $\bm{u}' \cdot \bm{u}'$ divided by $2\pi$ times the axial wavelength ($= 2\pi \cdot 2\pi / \kappa$), for consistency with the energy definition in \eqref{energy-integ}.

\begin{figure}
    \vspace{.1in}
    \centerline{\includegraphics[width=3.3in,keepaspectratio]{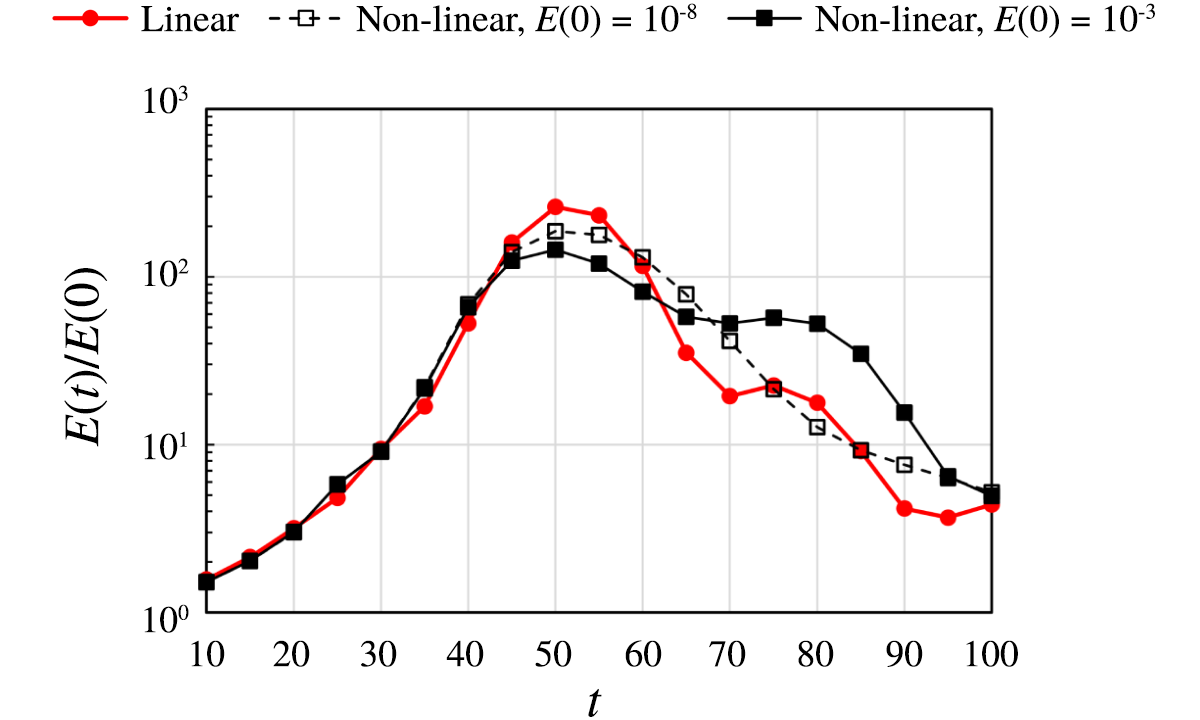}}
    \caption{Comparison of transient energy growth curves for the `linear' evolution case, calculated via \eqref{pert_evolve}, and for the `non-linear' evolution cases with initial perturbation energies of $10^{-8}$ and of $10^{-3}$, obtained from three-dimensional non-linear simulations. The initial perturbation is depicted in figure \ref{fig:vortinputz=0}.}
\label{fig:vortevolenergy}
\end{figure}

In figure \ref{fig:vortevolenergy}, three energy growth curves are plotted together for comparison. First, the energy growth curve in the linear evolution case peaks at $t=\tau=50$ with a maximum energy growth of $G = 3 \times 10^2$. This curve serves as an index of the linear process' prevalence during the early transient growth of vortices. Next, the energy growth curve in the non-linear evolution case with $E(0) = 10^{-8}$ aligns with the `linear index' curve. In this scenario, non-linear effects arise but remain minimal, showing a slight debilitation of maximum energy growth at $t=\tau=50$. However, it is unlikely that this change is entirely due to the non-linearity introduced to the system because, in comparison with \citet[][p. 55]{Mao2012}, such a drop in energy growth at the peak may also be attributed to viscous diffusion of the base flow over time. Lastly, the energy growth curve in the non-linear evolution case with $E(0) = 10^{-3}$ exhibits more pronounced debilitation at the peak. Compared to the previous case, this relatively high energy case demonstrates more intensification of the non-linearity, represented by a secondary energy hump around $t=80$ that is unpredicted by the linear case. Nevertheless, the general trend of the curve does not drift away from the linear prediction. Overall, the coherence in trend holds particularly well up to the maximum vortex growth period ($t < \tau = 50$), which suggests that the linearised transient growth framework remains effective in describing the early evolution of the original non-linear system. Beyond this period, the non-linearity intensifies, but the trend of decay remains persistent.

\begin{figure}
    \vspace{.1in}
    \centerline{\includegraphics[width=\textwidth,keepaspectratio]{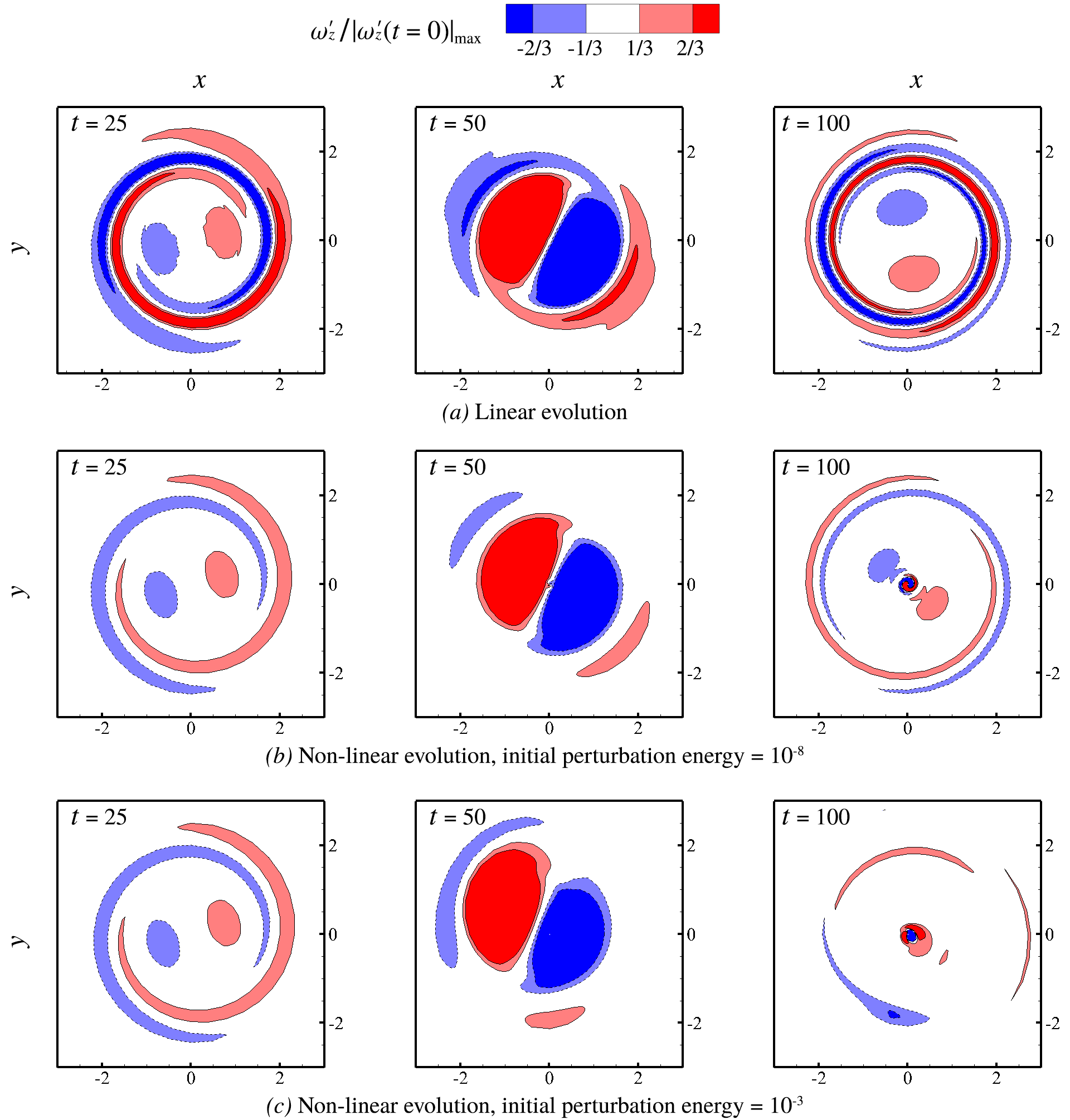}}
    \caption{Axial vorticity perturbation contours of the optimally perturbed vortex (refer to figure \ref{fig:vortinputz=0} for the initial perturbation) on the $z=0$ plane at $t=25$, $t=50$ and $t=100$: $(a)$ the `linear' evolution case where maximum energy growth is known to occur at $t=\tau=50$, $(b)$ the `non-linear' evolution case with initial perturbation energy of $10^{-8}$, $(c)$ the `non-linear' evolution case with initial perturbation energy of $10^{-3}$. The same contour style as figure \ref{fig:vortinputz=0} is used for all plots. Despite the presence and intensification of non-linearity with increasing perturbation energy over time, the `linear' process largely prevails the overall dynamics during early vortex growth in the original non-linear system.}
\label{fig:vortevolution}
\end{figure}

The prevalence of the linear process in the early-stage vortex growth becomes more evident when examining the evolution of the perturbation structure, as shown in figure \ref{fig:vortevolution}. Using the same contour style across the three cases discussed above, we present three snapshots of axial vorticity perturbation contours on the $z=0$ plane at $t=25$, $t=50$ and $t=100$ for each case. The structural coherence in vorticity perturbation between the linear and non-linear cases is apparent at $t=25$, representing the stage of rapid perturbation growth. At $t=50$, the optimal growth time, the perturbation structures remain largely coherent. However, in the non-linear evolution case with $E(0)=10^{-3}$, a weak breakdown from the $m=1$ symmetry becomes observable, indicating that other azimuthal wavenumbers rather than $m=1$ begin to gain non-negligible energy through higher-order energy transfer across different wavenumbers. By $t=100$, corresponding to the stage of asymptotic stabilisation, the perturbation structures no longer show strong resemblance. This is another indication of the non-linearity intensification resulting from prolonged vortex growth over time.

\begin{figure}
    \vspace{.1in}
    \centerline{\includegraphics[width=\textwidth,keepaspectratio]{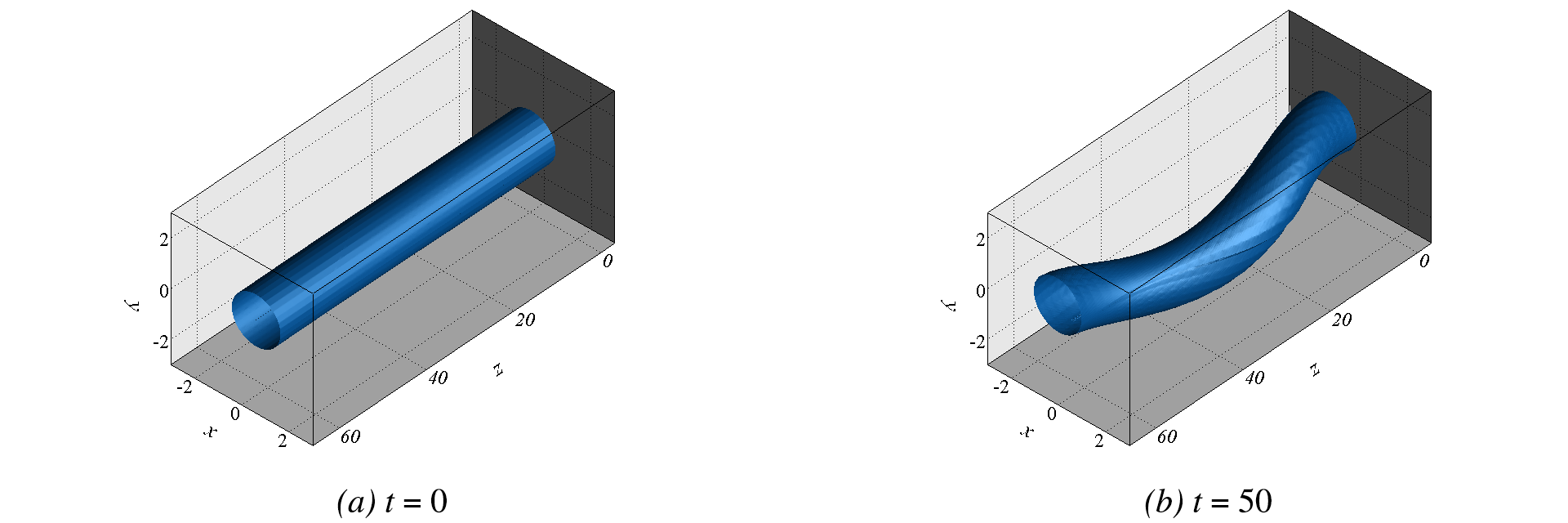}}
    \caption{Three-dimensional illustration of the $q$-vortex with a helical perturbation (see figure \ref{fig:vortinputz=0}), initiated with an energy level of $10^{-3}$: $(a)$ the initial core structure at $t=0$, and $(b)$ the most excited core structure at $t=\tau=50$. To detect the vortex core, the $\lambda_2$-isosurface at $\lambda_2 = -0.05$ is depicted \citep[see][]{Jeong1995}. The maximum displacement of the vortex centre comes up to the order of the core radius, comparable to experimental observations of vortex meandering amplitude \citep[see][]{Devenport1997, Bolle2021}.}
\label{fig:vortmeandering}
\end{figure}

Last but not least, we note that the simulation with initial perturbation energy of $10^{-3}$ results in substantial displacement of the vortex core, as shown in figure \ref{fig:vortmeandering}, notwithstanding the seemingly small initial energy level. The $\lambda_2$-isosurface at $\lambda_2 = -0.05$ is used to detect the vortex core \citep[see][]{Jeong1995}. The maximum displacement of the vortex centre in the simulation is nearly equal to the core radius, coinciding with experimentally observed meandering amplitudes of similar scale \citep[see][]{Devenport1997, Bolle2021}. Based on the rough figures for a large transport aircraft from \citet[][p. 259]{Fabre2004}, the characteristic scales in our formulation are $U_0 \approx 27~\mathrm{m/s}$ and $R_0 \approx 0.5~\mathrm{m}$. Using the density of air $\rho \approx 1~\mathrm{kg/m^3}$, the `dimensionless' energy of $10^{-3}$ translates to an `actual' kinetic energy of $(10^{-3}) \times 2\pi \rho (U_0^2/2) R_0^2 \approx 0.6~\mathrm{J/m}$ (`per metre' stands for axial unit length), which appears to be not exorbitant in practice. We believe this strengthens the practicability of the optimal transient growth process under consideration, along with the radially concentrated nature of the optimal perturbation structures (see figure \ref{fig:vortinputz=0}).

\section{Initiation of optimal transient growth}\label{sec:initiation}
\subsection{On a means of initiating optimal transient growth}\label{sec:onameans}
The transient evolution of the optimally perturbed $q$-vortex, as analysed in the previous section, unveils a promising way for significantly disturbing the vortex, even when the base vortex is known to be linearly stable. For this growth process to hold practical significance, an important question remains: By what means can such perturbations be initiated? In our analyses and simulations so far, the presence of perturbations has been presumed to be initially present along with the vortex supposedly in an undisturbed state. However, from a practical standpoint, there should be a means of introducing these perturbations, as they cannot spontaneously arise from the undisturbed flow — the $q$-vortex, which by itself is quasi-steady. Without a plausible initiation process, the optimal transient vortex growth process may remain purely theoretical.

Ambient turbulence might seem a compelling means of initiation, as often addressed in linear instability contexts \citep[e.g.,][]{Crow1976, Han2000}. However, unlike linear instability mechanisms — where perturbations are destined to be predominant in the limit of $t \rightarrow \infty$ due to the most unstable eigenmode's exponential growth — the transient growth process generally necessitates a specific (optimal) form of perturbation as input. Whether such a specific perturbation can spontaneously emerge from ambient turbulence, which is fundamentally stochastic and uncontrolled, has led to recurrent criticisms of optimal transient growth \citep[see][p. 235]{Fontane2008}. In the work by \citet{Fontane2008}, where the transient dynamics of vortices with stochastic forcing was examined, optimal perturbations were shown to be activated by noise-like forcing that is random in both space and time. This mitigates the aforementioned criticisms of optimal transient growth. Nonetheless, as the authors stated, it remains questionable whether such random isotropic forcing effectively represents turbulence in real conditions. The absence of clear universality in modeling turbulence poses a significant challenge in integrating ambient turbulence into this current problem.

Instead, we take an initiative in considering a different, non-stochastic means of initiating optimal transient growth: ice crystals (or particles). The presence of such particles in real-world scenarios is ascertained through contrail observations. Contrail formation primarily starts with jet exhaust plumes produced by aircraft engines, which contain particulate matter that eventually acts as condensation nuclei \citep{Karcher2018}. However, our interest does not lie in this very initial contrail stage during jet plume development, when the vortex roll-up process is still ongoing. According to early experimental findings by \citet{El-Ramly1977}, there is no appreciable influence of the engine exhaust on altering the rolled-up structure at this stage. As a matter of course, our attention is directed toward the later stage involving a fully formed wake vortex, where interactions between the vortex and ice crystals becomes manifest.

During the stage when ice crystals interact with the wake vortex, individual particle sizes reach a few microns, or approximately 1 to 5 microns \citep{Karcher1996, Paoli2005, Naiman2011, Voigt2011, Karcher2018}. It is noted that, according to \citet{Voigt2011}, actual contrail samples exhibited particle sizes ranging from 0.39 to 17.7 microns, with an effective radius of 2.9 microns. Given the substantial size difference between the vortex scale (measured in metres) and the ice crystals (microns), thes particles are often treated as flow tracers, i.e., with no backward influence on the carrier fluid \citep{Paoli2005, Naiman2011}. This assumption can reasonably simplify the particle-flow dynamics. However, when considering a large particle number density, reportedly ranging $10^{9}$ to $10^{11}$ per cubic metre \citep{Paoli2004, Paoli2005}, combined with the density ratio of ice to air (approximately $10^{3}$), their bulk impact may not be simply ignored. In the most optimistic estimation based on the given figures (a particle radius of 5 microns and a particle number density of $10^{11}$ per cubic metre), the upper limit of the particle mass fraction could fall between $10^{-2}$ and $10^{-1}$, which, although representing one extreme end, is not preposterous. This mass fraction is substantial enough to initiate perturbations ultimately evolving into significant disturbances (recall \S \ref{sec:nonlinearimpacts}).

When considering ice crystals, a primary emphasis has typically been on their microphysical growth, typically using an ice microphysics model alongside a flow solver \citep[e.g.,][]{Lewellen2001, Paoli2004, Paoli2005, Naiman2011}. We exclude the context of microphysical growth in this study and, instead, direct our attention to two-way coupling, specifically through drag momentum exchange. This direction aligns with our essential goal of exploring whether particle-induced drag can initiate optimal transient growth. 

Inspired by natural contrail formation, our analysis in this section seeks to shed light on how particles’ backward influence can act as a `controlled' perturbation mechanism to trigger optimal transient growth. Unlike ambient turbulence, which is stochastic and uncontrollable, particles may offer the potential for a more deliberate approach. By strategically releasing particles around the vortex periphery, e.g., adjusting their distance from the vortex or timing their ejection period, it is possible to introduce targeted perturbations that invoke the optimal structures identified for transient growth.


Recent studies by \citet{Shuai2022A} and \citet{Shuai2022} indicate that weakly inertial particles within a vortex, under two-way coupled conditions, can trigger instabilities and expedite the process of vortex decay. While the initial particle distributions considered in these studies — where particles are loaded either throughout the entire domain or inside the vortex core — are not directly applicable to our case, where particles are at the periphery of the vortex core and interact with the vortex, these studies support the underlying concept that even a dilute amount of particles can meaningfully influence the surrounding vortex. By considering particles, our study aims at broadening the understanding of transient growth dynamics in a controlled manner, providing insights into potential applications of particle-driven perturbations.

\subsection{Two-way coupled equations for vortex-particle interaction}\label{sec:two-waycoupled}
To simulate the initiation process leading to transient vortex growth via particle drag, it is necessary to introduce additional parameters and variables in order to establish the equations governing particle motion. Additionally, a coupling term must be added to the momentum equation of fluid motion in \eqref{nonlingoveqn} to complete a two-way coupled formulation. In this study, we consider dispersed ice crystals, whose density is approximately $10^3$ times that of the surrounding fluid (air). We define the ratio of particle density $\rho_{p}$ to fluid density $\rho$ as a new dimensionless parameter, denoted by $\vartheta$ ($\equiv \rho_{p} / \rho$). For subsequent calculations, we set $\vartheta$ to a constant value of $10^3$.

In this study, we employ the Eulerian approach adopting the fast equilibrium approximation proposed by \citet{Ferry2001}. This method treats the set of particles as a continuum, allowing the flow-particle system to behave like a two-phase flow. Given the high particle number density, we opt against Lagrangian approaches that track all individual particles \citep[e.g.,][]{Paoli2004, Naiman2011, Shuai2022}. Thanks to the relatively moderate computational cost, we believe that the Eulerian approach is favourable for future scale-up simulations involving two or more vortices to explore secondary vortex evolution. Moreover, our focus on the `bulk' influence of particles on the surrounding vortex, rather than individual particle statistics, justifies the continuum treatment of particles.

The following two variables now represent the particles in the dispersed phase form: the particle velocity field, $\bm{u}_p (r,\phi,z, t)$ and the particle volume fraction, $c (r,\phi,z, t)$. The fast equilibrium approximation enables explicit evaluation of $\bm{u}_p$ in terms of the fluid velocity field, $\bm{u}$. Using the Maxey-Riley equation with the added mass effect \citep{Maxey1983, Auton1988}, $\bm{u}_p$ can be reduced and the resulting two-way coupled equations \citep[see][p. 1221]{Ferry2001} are
\begin{equation}
\frac{\partial \bm{u}}{\partial t} = - \bm{\nabla} \varphi + \bm{u} \times \bm{\omega} + \frac{1}{\Rey} {\nabla}^2 \bm{u} - (\vartheta - 1)c \frac{D\bm{u}}{Dt} ~~~~~\text{with}~~\bm{\nabla} \cdot \bm{u}=0,
    \label{twowaymomeqn}
\end{equation}
and
\begin{equation}
    \frac{\partial c}{\partial t} = - \bm{u} \cdot \nabla c +  \frac{2 {Stk} (\vartheta - 1)}{2\vartheta + 1} \nabla \cdot \left( c \frac{D\bm{u}}{Dt} \right),
    \label{twowaypvoleqn}
\end{equation}
where $D/Dt \equiv \partial/\partial t + \bm{u} \cdot \nabla$ is the material derivative with respect to the fluid phase and $Stk$ is the Stokes number, i.e., the dimensionless particle relaxation time normalised by $R_0/U_0$. For calculations, $Stk$ is set to $10^{-5}$ to maintain compatibility with the fast equilibrium approximation as well as practical conditions \citep{Karcher1996}. 
The discretisation and time-integration procedures for \eqref{twowaymomeqn} and \eqref{twowaypvoleqn} are not different from those used in the previous pure vortex cases, as detailed in Appendix \ref{appB}. Compared to \eqref{nonlingoveqn}, it is found that the last term in \eqref{twowaymomeqn} accounts for the particle drag, with its magnitude dependent on the order of $c$.

\subsection{Initial particle distribution}
As discussed in recent studies on vortex-particle interactions \citep{Shuai2022A, Shuai2022}, the observable effects of these interactions depend on the initial particle distribution. To explore whether particle drag can initiate optimal transient growth in the vortex, it is necessary to establish an effective initial distribution of the particles. In the following discussion, we first emphasise that the particles' influence is limited to generating only a weak disturbance within the vortex system, under the assumption that $(\vartheta - 1)c$ remains significantly less than order unity, comparable to the perturbation order.

To comprehend how the particles induce perturbations in the carrier fluid, we reorganise the coupled momentum equation in \eqref{twowaymomeqn}, yielding
\begin{equation}
    \left( 1 + (\vartheta - 1 ) c \right) \frac{D \bm{u}}{D t} = - \bm{\nabla} p + \frac{1}{\Rey} {\nabla}^2 \bm{u},
    \label{twowaymomeqn-revised}
\end{equation}
which is in fact the original formulation provided by \citet{Ferry2001}. In this form, it is evident that the combined motion of the two phases resembles a single-phase flow with slight density variations, as characterised by the factor $(1 + (\vartheta - 1) c)$. This effect is understood as a consequence of the dispersed phase absorbing momentum from the fluid phase. Loosely speaking, if the fluid accelerates, the particles' presence retards the fluid's acceleration, resulting in a negative perturbation velocity, and vice versa. 

Given that the perturbation velocity is induced by non-zero $c$, we derive the perturbation momentum equation from the decompositions of velocity and pressure in \eqref{twowaymomeqn-revised} (i.e., $\bm{u} = \overline{\bm{u}} + \bm{u}'$ and $p = \overline{p} + p'$), with the assumption that $(\vartheta - 1 ) c$ is comparable to the perturbation (prime) order, e.g., $O(\varepsilon)$. Since zero particle concentration implies no perturbation, the mean quantity (overscore) equation is obtained by setting $c=0$ in \eqref{twowaymomeqn-revised}, yielding ${D \bm{\overline{u}}}/{Dt} = - \bm{\nabla}\, \overline{p} + \nabla^2{\overline{\bm{u}}}/{\Rey}$. Subtracting this from \eqref{twowaymomeqn-revised} and neglecting terms of order higher than $O(\varepsilon)$, we obtain
\begin{equation}
    (\vartheta - 1 ) c \left[ \frac{\partial \overline{\bm{u}}}{\partial t} + (\overline{\bm{u}} \cdot \nabla )\,\overline{\bm{u}} \right] + \frac{\partial \bm{u}'}{\partial t} + (\overline{\bm{u}} \cdot \nabla )\,\bm{u}'= - \bm{\nabla} p' + \frac{1}{\Rey} {\nabla}^2 \bm{u}'.
    \label{twowaymomeqn-revised-decomposed}
\end{equation}
Our goal is to estimate the initial particle distribution $c(t=0)$, denoted $c_0$, that effectively perturbs the `undisturbed' vortex towards optimal transient growth, i.e., $\bm{u}'(t=0) = 0$. Under this condition, \eqref{twowaymomeqn-revised-decomposed} reduces to the initial perturbation growth formula: 
\begin{equation}
    \frac{\partial \bm{u}'}{\partial t}\Big\rvert_{\, t=0} = - \bm{\nabla} p' (t=0) -(\vartheta - 1 ) c_0 \frac{D\overline{\bm{u}}}{Dt}\Big\rvert_{\, t=0},
    \label{twowaymomeqn-revised-decomposed-t0}
\end{equation}
which describes the growth of $\bm{u}'$ over a brief initial period, e.g., $0 \le t < \delta t \ll 1$. After a brief advance of time, the spatial structure of $\bm{u}'$ is primarily attributed to that of $-(\vartheta - 1 ) c_0D\overline{\bm{u}}/Dt\rvert_{\, t=0}$ (initial particle drag) while $- \bm{\nabla} p' (t=0)$ serves only to maintain the divergence-free constraint. In other words, we expect $\bm{u'}(t=\delta t) \approx \delta t [ - (\vartheta - 1 ) c_0D\overline{\bm{u}}/Dt\rvert_{\, t=0} ]$ with a minor adjustment for the field to be solenoidal (for details of the adjustment, see Appendix \ref{appC}).

Suppose that we aim to initiate a known optimal velocity perturbation $\Breve{\bm{u}}'$. If we identify $c_0$ such that $-(\vartheta - 1 ) c_0D\overline{\bm{u}}/Dt\rvert_{\, t=0}$ matches (a positive constant multiple of) $\Breve{\bm{u}}'$, then we can expect $\bm{u}'$ to exhibit this perturbation form. However, the existence of such $c_0$ is rare, since this matching requires solving three component equations whereas $c_0$ is the only unknown. To circumvent this overdetermination issue, we inevitably focus on the most critical component among them. By choosing the azimuthal component, the problem reduces to finding $c_0$ such that
\begin{equation}
    -(\vartheta - 1) c_0 { \bigg( \frac{D\overline{\bm{u}}}{Dt} \bigg)_\phi \, \Big\rvert_{\, t=0}} 
    = { C \Breve{u}'_\phi }, 
    \label{initialdistopt}
\end{equation}
where $C$ is an arbitrary positive constant, which can be used to scale $c_0$ when solving \eqref{twowaymomeqn} and \eqref{twowaypvoleqn} for different levels of particle volumetric loading. Arranging the terms based on the fact that $(D\overline{\bm{u}}/Dt)_\phi \rvert_{t=0} = -4r e^{-r^2} / \Rey$ for the `undisturbed' $q$-vortex profile, we can further simplify this relation to $c_0 \propto \Breve{u}'_\phi/(r e^{-r^2})$.

The proposed $c_0$ has two serious limitations that restrict its practical applicability. Nevertheless, we affirm that it remains sufficiently effective for initiate optimal transient growth. First, the radial and axial components of perturbation velocity are excluded in this derivation. This is justifiable since the azimuthal component of velocity perturbation is generally dominant during the initial stages of transient growth (see the left panels in figure \ref{fig:pert_struct}). Second, the particle volume fraction cannot take on negative values. The continuity of the fluid may alleviate this issue; in a local sense, a deficit (or surplus) in speed within the particle-laden fluid must be offset by a corresponding gain (or loss) of speed in the circumferential particle-free fluid. As a result, we set $c_0$ to locally zero when its calculated value from \eqref{initialdistopt} is locally negative.

\begin{figure}
    \vspace{.1in}
    \centerline{\includegraphics[width=\textwidth,keepaspectratio]{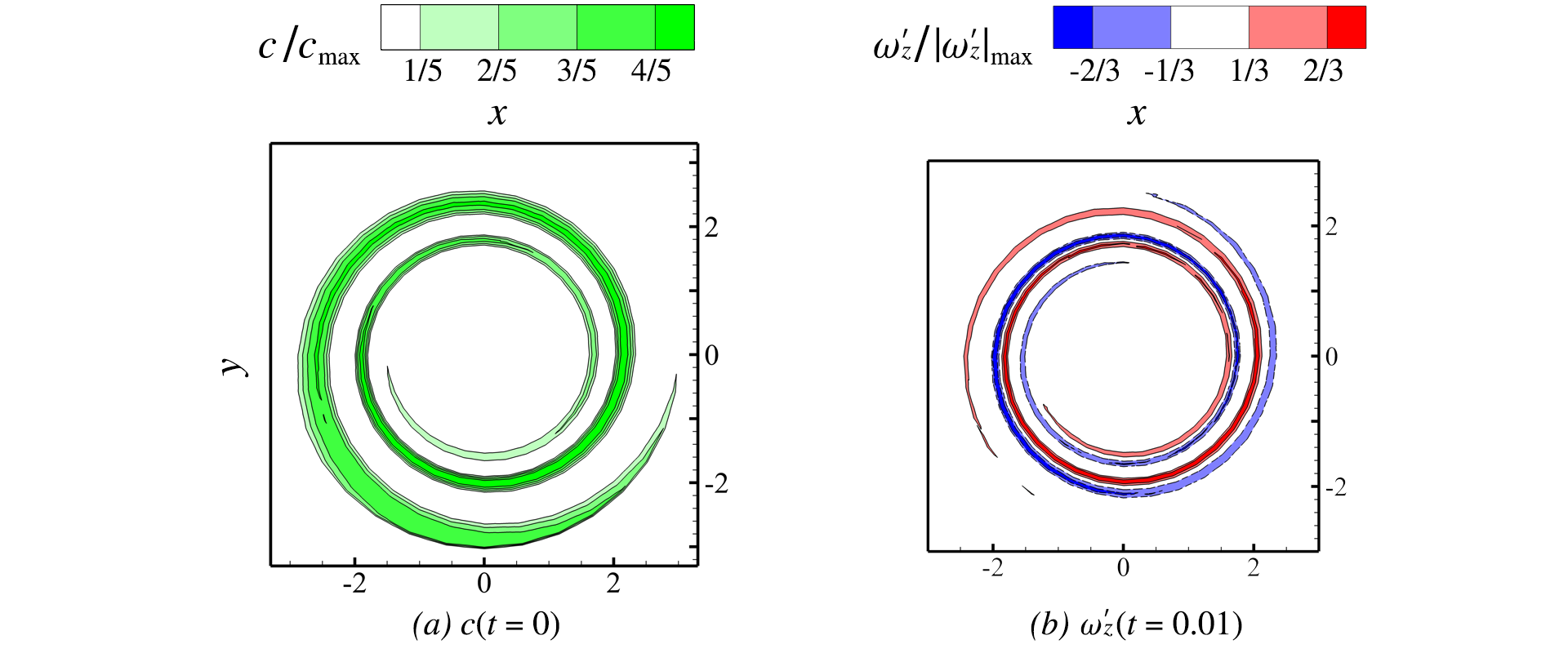}}
    \caption{Initiation of optimal transient growth via inertial particles: $(a)$ the initial particle volume fraction contour on the $z=0$ plane, in the pursuit of initiating the perturbation discussed in \S \ref{sec:nonlinearimpacts} (see figure \ref{fig:vortinputz=0}), and $(b)$ the axial vorticity perturbation contour on the $z=0$ plane after a brief advancement in time ($t=0.01$) in the two-way coupled vortex-particle simulation, solving \eqref{twowaymomeqn} and \eqref{twowaypvoleqn} with an initial particle volumetric loading level of $c_{\max}  = 10^{-6}$. In the right panel, the same contour style as figure \ref{fig:vortinputz=0} is used and $|\omega_z' (t=0.01)|_{\max} = 5.80 \times 10^{-6}$.}
\label{fig:partinputz=0}
\end{figure}

For comparison's sake, we revisit the optimal perturbation case discussed in \S \ref{sec:nonlinearimpacts}, where $(m,\,\kappa) = (1,\,0.1)$ with $\tau = 50$. Figure \ref{fig:partinputz=0} shows the initial particle volume fraction calculated using the proposed estimation, along with the axial vorticity after a brief advancement in time ($t=0.01$), resulting from the two-way interactions between the particles and the vortex with an initial particle volumetric loading level of $c_{\max} = 10^{-6}$. The computation of perturbation velocity fields in two-way coupled vortex-particle simulations follows the same approach as described in \eqref{nonlinperteval}, except that what determines the perturbation intensity is now the particle volumetric loading level $c_{\max}$. Despite the previously mentioned limitations, the resulting perturbation effectively resembles the intended perturbation input (see figure \ref{fig:vortinputz=0}) needed for optimal transient growth. This observation provides \textit{a posteriori} validation of \eqref{initialdistopt}.

\subsection{Particle-initiated transient growth}
Now that the vortex evolution is governed by \eqref{twowaymomeqn}, where the term $(\vartheta - 1) c (D {\bm{u}}/Dt)$ serves as an additional forcing term, the overall perturbation dynamics are influenced not only by the transient growth process but also by the continual interaction between the particles and the vortex flow over time \citep[see also a simliar discussion in][p. 249]{Fontane2008}. In what follows, we substantiate particle-initiated transient growth by identifying some notable indications of transient growth over short time intervals in the vortex-particle system, using the initial particle distribution shown in figure \ref{fig:partinputz=0}.

\begin{figure}
    \vspace{.1in}
    \centerline{\includegraphics[width=\textwidth,keepaspectratio]{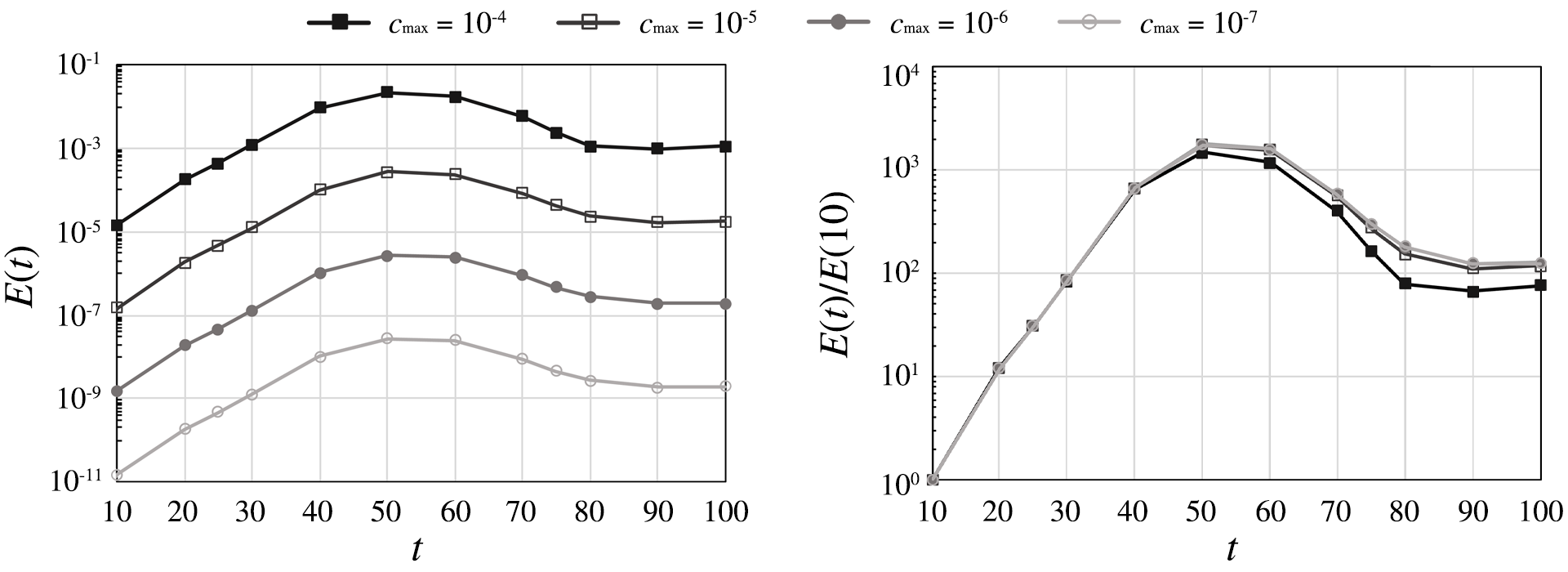}}
    \caption{(Left) Temporal changes in perturbation energy for vortex-particle interactions (see figure \ref{fig:partinputz=0} for the initial setup), evaluated at various levels of particle volumetric loading: $c_{\max} = 10^{-4}$, $10^{-5}$, $10^{-6}$ and $10^{-7}$, and (right) the same data, normalised by $E(10)$, to facilitate comparison of energy amplification across cases. Note that $E(10)$ is used for normalisation because $E(t)/E(0)$ is undefined here ($E(0) = 0$).}
\label{fig:partenergygrowth}
\end{figure}

Temporal changes in perturbation energy are displayed in figure\ref{fig:partenergygrowth} for four different levels of particle volumetric loading: $c_{\max} = 10^{-4}$, $10^{-5}$, $10^{-6}$, and $10^{-7}$. The case of $c_{\max} = 10^{-4}$ is considered to be the upper limit where $(\vartheta - 1) c$ remains significantly less than order unity (n.b., $\vartheta = 10^3$). All energy curves exhibit a nearly identical trend, suggesting that similar dynamics govern every case. In the right panel, the data are normalised by the perturbation energy at $t=10$ for each case (note that $E(0)$ is zero and the energy growth used, $E(t)/E(0)$, becomes undefined here), allowing for a comparison of energy amplification across cases. Overall, energy amplification levels off at around $10^2$ times $E(10)$, due to the long-term response of the vortex to the particle interactions. Arguably, amplification beyond this level should be attributable to the transient growth process, as evidenced by the energy amplification `hump' up to $t = 80$. Notably, the peak of this hump at $t=50$ coincides with the optimal transient growth period. $\tau = 50$, as intended to induce, which further supports our argument.

\begin{figure}
    \vspace{.1in}
    \centerline{\includegraphics[width=\textwidth,keepaspectratio]{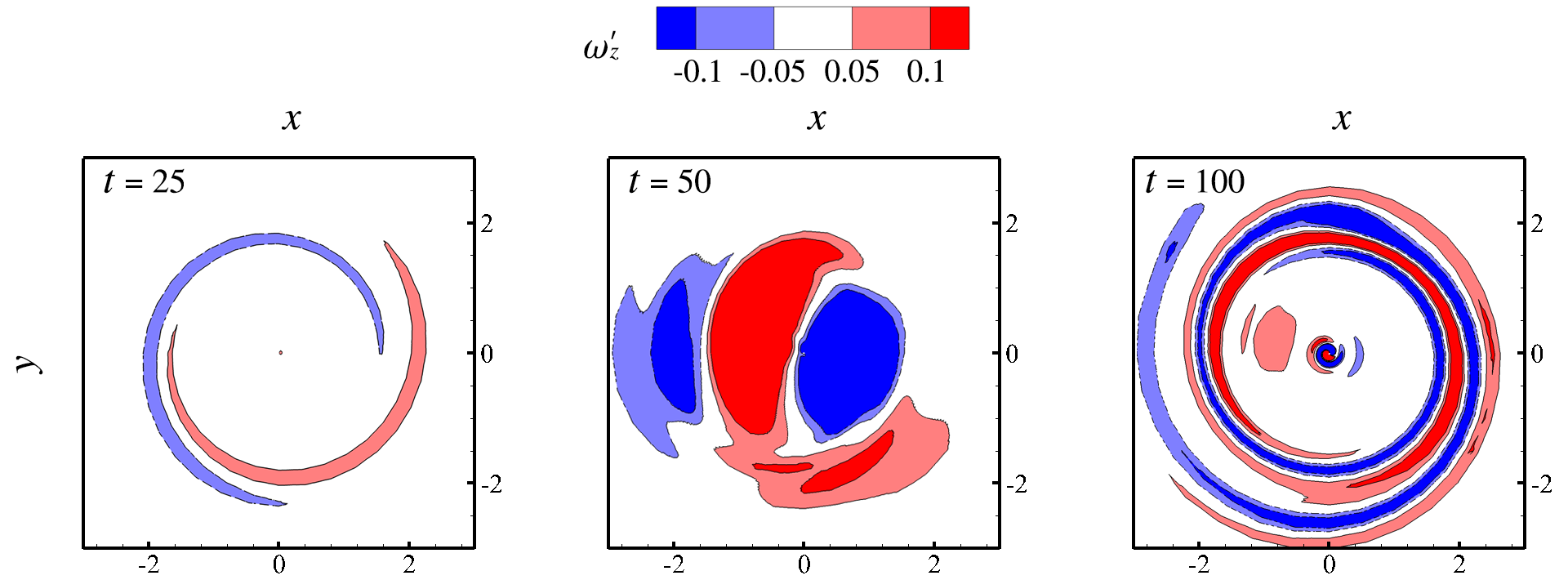}}
    \caption{Axial vorticity perturbation contours of the vortex interacting with particles initially distributed around the periphery (see figure \ref{fig:partinputz=0} for the initial particle distribution) on the $z=0$ plane at $t=25$, $t=50$ and $t=100$. Here, depicted is the case with $c_{\max} = 10^{-4}$.}
\label{fig:partvortevolution}
\end{figure}

Another indication of the particle-initiated transient growth is observed in the evolution of the perturbation structure. In figure \ref{fig:partvortevolution}, axial vorticity perturbation contours on the $z=0$ plane at $t=25$, $t=50$ and $t=100$ in the vortex-particle simulation with $c_{\max} = 10^{-4}$ are shown. These snapshots are compared with those of the optimally perturbed non-linear vortex growth in figure \ref{fig:vortevolution}$(c)$. The continual vortex-particle interactions produce structural discrepancies in the perturbation; this is clearly discernible at $t=100$, where strong spiraling arms form at the periphery as a result of prolonged drag momentum exchange. Nonetheless, during the early-stage perturbation growth up to $t=50$, key features characterising the optimal transient growth process are observed, such as the appearance of two weak spiraling arms at the periphery of the core at $t=25$. Most notably, at the time of maximum energy growth ($t=50$), perturbation energy transfer from the periphery to the core — the iconic feature of the optimal transient growth process for $m=1$ — is identifiable. We believe that this provides plausible evidence that near-optimal transient growth takes place through vortex-particle interactions.

\begin{figure}
    \vspace{.1in}
    \centerline{\includegraphics[width=\textwidth,keepaspectratio]{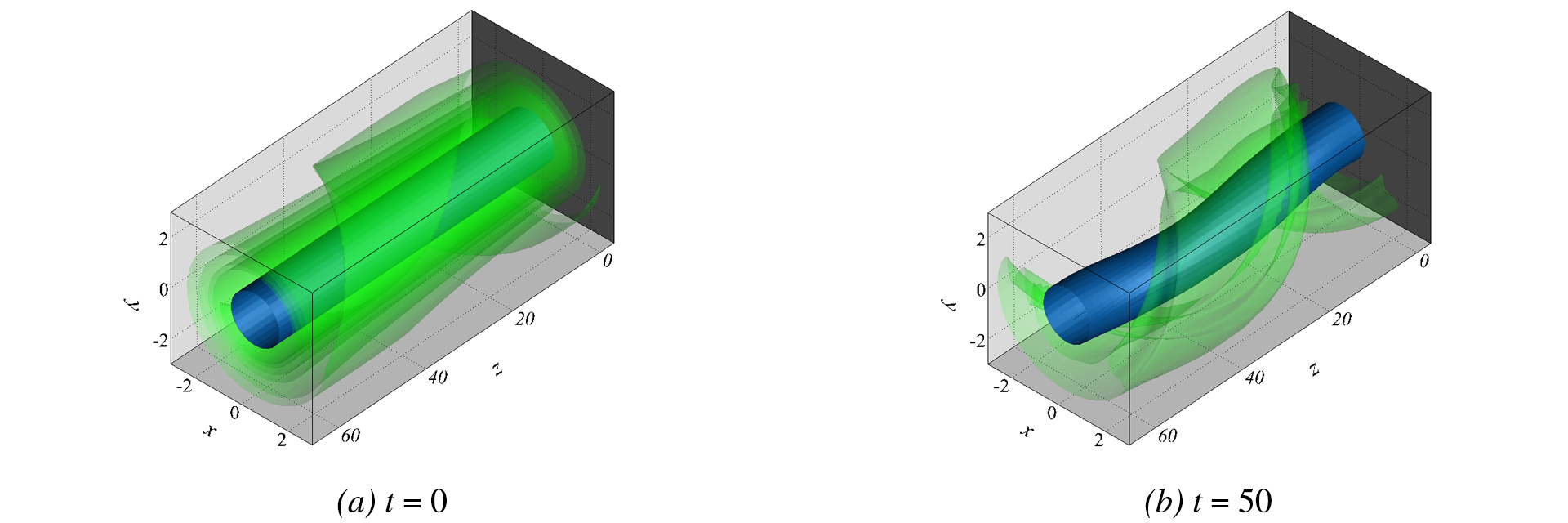}}
    \caption{Three-dimensional illustration of the $q$-vortex interacting with the peripherally located particles (see figure \ref{fig:partinputz=0}) of the initial particle volumetric loading level of $c_{\max} = 10^{-4}$: $(a)$ $t=0$, and $(b)$ $t=50$. The vortex core is detected using the $\lambda_2$-isosurface where $\lambda_2 = -0.05$, as with figure \ref{fig:vortmeandering}. The green isosurface of $c = 2 \times 10^{-5}$ (20 \% of $c_{\max}$) is drawn together to visualise the particle distribution at each time.}
\label{fig:partvort3d}
\end{figure}

Lastly, we report the transient development of the particle distribution in association with the vortex's transient growth. In figure \ref{fig:partvort3d}, the interaction between the vortex and particles for the case $c_{\max} = 10^{-4}$ is visualised, using the $\lambda_2$-isosurface at $\lambda_2 = -0.05$ for the vortex core and the $c$-isosurface representing 20 \% of $c_{\max}$ for the particles. As the vortex evolves from its unperturbed state at $t=0$ to its most excited state at $t=50$, the particles show reduced dispersion, forming a coherent helical structure that envelops the vortex core. However, this coherence is evanescent and dissipates rapidly after the peak perturbation growth at $t=50$. The physical significance of this temporary coherence increase, as well as whether this is exclusive to the $m=1$ case, warrants further investigation. For now, it is noteworthy that the physical phenomenon linked to the current example, vortex meandering, is known to increase system `orderliness' (i.e., reduce the number of dynamically active proper orthogonal decomposition (POD) modes), as reported by \citet{Bolle2021}; the coherence of the particles during transient vortex growth may be associated with this tendency.

\section{Conclusion}\label{sec:conclusion}
In this study, we investigated the transient dynamics of a wake vortex using a spectral method specifically devised for a radially unbounded domain. Our findings confirmed that the primary contributor to optimal transient growth among continuous eigenmode families is the viscous critical-layer eigenmode family, rather than the potential eigenmode family. Additionally, we explored the role of inertial particles located at the periphery of a vortex, inspired by ice crystals forming contrails in real-world scenarios, as a significant means of initiating optimal transient growth via drag momentum exchange — an effect that is often overlooked.

Using the spectral method for an unbounded domain, developed with mapped Legendre functions as basis functions \citep[see][]{Lee2023}, we numerically analysed the transient growth process of the $q$-vortex when slightly disturbed by a perturbation formed as a sum of well-resolved eigenmodes. Unlike the conventional spectral method involving Chebyshev polynomials with domain truncation, our method is not susceptible to critical issues that adversely impact numerical irrelevant in transient growth analysis. These issues include the excessive generation of unnecessary (non-regular or spurious) eigenmode families and the lack of clear distinction between the viscous critical-layer spectrum and the potential spectrum, due to the incomplete approximation of the unbounded domain. By addressing these problems proactively, our method offers greater flexibility for adjusting numerical resolution through the map parameter $L$.

Following the typical transient growth formalism, we treated perturbations as a sum of eigenmodes and subsequently explored which family of eigenmodes primarily contributes to optimal perturbations for achieving maximum transient growth. The important behaviour of short-term perturbation energy growth is associated with continuously varying eigenmodes, grounded upon the non-normality of the linearised Navier-Stokes operator. While \citet{Mao2012} demonstrated the dominance of continuous eigenmodes in optimal perturbations for transient growth in a wake vortex, their study did not further categorise the continuous eigenmodes, particularly those belonging to the viscous critical-layer eigenmode family. 

Through an analysis of sub-eigenspaces, each spanned by a distinct eigenmode family, it was corroborated that optimal transient growth is primarily attributed to the viscous critical-layer eigenmodes. This finding provides a better alignment with the theoretical foundation of transient growth with the critical layer analysis, rather than with the wave packet pseudomode analysis, supporing the argument that inviscid continuous spectrum (CS) transients drive vortex growth over short time intervals \citep{Heaton2007Optimal}. It also refines the focus when exploring the continuous spectra, as the viscous critical-layer spectrum accounts solely for continuous curves adjacent to the discrete spectrum, whereas the remaining continuous spectrum (potential spectrum) covers a much wider area in the left half of the complex eigenvalue plane.

The energy growth curves and associated optimal perturbation structures obtained in the current analysis, for both axisymmetric ($m=0$) and helical ($m=1$) cases with axial wavenumbers $\kappa$ of order unity or less, were consistent with previous studies. We identified the general responses during optimal transient growth: for $m=0$ a transition from azimuthal velocity to other components, forming consistent ring-like streaks; and for $m=1$, a shift from swirling velocity layers outside the vortex core to substantial transverse motion within the core. These results align with earlier research on vortex transient growth, such as by \citet{Pradeep2006, Fontane2008, Mao2012}.  In non-linear simulations of the $q$-vortex, initiated with an $(m,\,\kappa)=(1,\,0.1)$ optimal perturbation over the optimal growth period $\tau=50$, transient growth dynamics persisted notably up to the time of maximum energy growth ($t \le \tau = 50$). The linear prediction of transient growth effectively captures early development in the original non-linear vortex system, even when perturbation growth becomes comparable to the base flow order and causes visible distortion (see figures \ref{fig:vortevolenergy} and \ref{fig:vortmeandering}.)

Lastly, we discussed the initiation process of transient growth — generating perturbations through physical interactions rather than assuming their presence from the start — and evaluated its validity. While ambient turbulence may offer a potential mechanism for initiating transient growth, its inherent complexity, involving a broad range of scales and stochastic dynamics, complicates precise modelling. To bypass these complexities, we focused on vortex-particle interactions inspired by ice crystals (or contrails) associated with aircraft wake vortices.

Despite the small size of individual particles, which often leads to the assumption that their influence on the flow is negligible, their bulk inertial effect, combined with high particle number density, can have a significant impact. By enabling two-way coupling between particles and the vortex flow via drag momentum exchange, we conducted vortex-particle simulations, with particles initially distributed at the periphery of the vortex core to initiate the optimal perturbation for $(m,\,\kappa)=(1,\,0.1)$ studied earlier. The simulations showed clear signs of optimal transient growth during the continual vortex-particle interactions, including notable energy amplification peaking at $t=\tau=50$ and the transfer of perturbation energy from the periphery to the core.

The present study underscores the crucial role of the optimal transient growth process of a single vortex over short time intervals, initially structured by critical-layer eigenmodes. The initiation of transient growth via particle drag not only demonstrates the practicability of this transient growth process but also highlights the susceptibility of vortex motion, even to physical interactions that are often overlooked for simplicity or perceived insignificance. As \citet{Fontane2008} suggested in their study on vortex transient growth study with stochastic forcing, the transient growth process may be active regardless of the specific characteristics or dynamics of the perturbations; particle drag could serve as one of these activators. 

Even though our motivation of considering particles was founded upon contrails in real-world scenarios, the use of particles to perturb a vortex can also be justified for purposes beyond understanding its nature, such as actively controlling the wake vortex system to hasten its destabilisation, potentially through deliberate injection of inertial particles. Leveraging the same numerical scheme (see Appendix \ref{appB}), we expect to extend the scope to our analysis to vortex pairs or multi-vortex systems to explore whether the transient growth of individual vortices contributes to faster onset of practical and established instabilities in aircraft wake vortices, such as the Crow instability, facilitating more expeditous vortex destabilisation.

\backsection[Acknowledgements]{We would like to thank Dr. Joseph A. Barranco (San Francisco State University) for his valuable insights about particle-laden flow simulations using the fast equilibirum assumption, and Jinge Wang (University of California, Berkeley) for engaging discussions on wake vortex instabilities.}

\backsection[Funding]{This research received no specific grant from any funding agency, commercial or not-for-profit sectors.}

\backsection[Declaration of interests]{The authors report no conflict of interest.}

\appendix
\section{Numerical integration for energy calculation}\label{appA}
Consider the following definite integral $I$ of an arbitrary scalar $f$ in terms of $r$:
\begin{equation}
    I(f) \equiv \int_{0}^{\infty} {f^*(r) f(r) rdr}.
    \label{scalar-integ}
\end{equation}
It is assumed that $f$ decays sufficiently fast as $r \rightarrow \infty$ so that $I(f)$ is well-defined. Using a change of variables from $r \in \left[ 0,\infty \right)$ to a new variable $\zeta \in [-1,1)$ via
\begin{equation}
    \zeta \equiv \frac{r^2 - L^2}{r^2 + L^2},
    \label{mapparam}
\end{equation}
where $L>0$ is the map parameter, we transform \eqref{scalar-integ} into a new form as follows:
\begin{equation}
    I(f)= \int_{-1}^{1} {f^* \left( L \sqrt{\frac{1+\zeta}{1-\zeta}} \right) f \left( L \sqrt{\frac{1+\zeta}{1-\zeta}} \right) \frac{L^2}{(1-\zeta)^2} d\zeta}.
    \label{scalar-integ-new}
\end{equation}
Applying the Gauss-Legendre quadrature rule as employed by \citet[p. 13]{Lee2023}, the numerical form of \eqref{scalar-integ-new} becomes
\begin{equation}
    I(f) \simeq \sum_{j=1}^{N} f^*(r_j) \frac{L^2 \varpi_j}{(1 - \zeta_j)^2} f(r_j) ,
    \label{scalar-integ-numeric}
\end{equation}
where $\zeta_j$ and $\varpi_j$ are the $j$th abscissa and weight of the Gauss-Legendre quadrature rule for degree $N$ $(j = 1,~2,\cdots,~N)$, and $r_j \equiv L \sqrt{(1 + \zeta_j)/(1 - \zeta_j)}$ is the $j$th radial collocation point. Note that \eqref{scalar-integ-numeric} can be expressed as $\bm{f}^* \mathsfbi{M}^{\ddag} \bm{f}$ if we define $\bm{f}$ as the discretised version of $f$ in physical space, i.e., $\bm{f} \equiv \left( f(r_1),~\cdots,~f(r_N) \right)$ and $\mathsfbi{M}^{\ddag}$ as
\begin{equation}
    \mathsfbi{M}^{\ddag} \equiv \text{diag} \left( \frac{L^2 \varpi_1}{(1 - \zeta_1)^2},~\cdots,~\frac{L^2 \varpi_N}{(1 - \zeta_N)^2} \right).
    \label{M-block}
\end{equation}

Finally, the energy in \eqref{energy-integ}, which is equal to $I(\Tilde{\bm{u}}_r) + I(\Tilde{\bm{u}}_\phi) + I(\Tilde{\bm{u}}_z)$, can be numerically calculated as follows:
\begin{equation}
    E(\Tilde{\bm{u}}) = \underbrace{\left(\begin{array}{c;{2pt/1pt}c;{2pt/1pt}c}\Tilde{\bm{\upsilon}}_r^* & \Tilde{\bm{\upsilon}}_\phi^* & \Tilde{\bm{\upsilon}}_z^*\end{array}\right)}_{\Tilde{\bm{\upsilon}}^*} \underbrace{\left( \begin{array}{c;{2pt/1pt}c;{2pt/1pt}c} \mathsfbi{M}^{\ddag} & \mathsfbi{0} & \mathsfbi{0} \\[2pt] \hdashline[2pt/1pt] \mathsfbi{0} & \mathsfbi{M}^{\ddag} & \mathsfbi{0} \\[2pt] \hdashline[2pt/1pt] \mathsfbi{0} & \mathsfbi{0} & \mathsfbi{M}^{\ddag} \end{array} \right)}_{\mathsfbi{M}} \underbrace{\left(\begin{array}{c} \Tilde{\bm{\upsilon}}_r \\[2pt] \hdashline[2pt/1pt] \Tilde{\bm{\upsilon}}_\phi \\[2pt] \hdashline[2pt/1pt] \Tilde{\bm{\upsilon}}_z \end{array}\right)}_{\Tilde{\bm{\upsilon}}}.
    \label{energy-integ-matrix}
\end{equation}
This constitutes the formation of $\mathsfbi{M}$ representing the matrix calculation for energy in \eqref{energy-matrix}. 

\section{Numerical setup for non-linear simulations}\label{appB}
To discretise the radially unbounded domain considered in this paper, especially in three dimensions for non-linear simulations, we employ a pseudo-spectral method based on the mapped Legendre spectral collocation method. The method assumes that an arbitrary scalar field (or a component of an arbitrary vector field) that decays rapidly and harmonically in $r$, say, $f$, is expanded as follows:
\begin{equation}
    f(r,\phi,z,t) = \sum_{k=-\infty}^{\infty} \sum_{m=-\infty}^{\infty} \sum_{n=|m|}^{\infty}   f_{n}^{mk}(t) P_{L_{n}}^{m} (r) e^{im\phi} e^{i k \frac{2 \pi }{Z}z},
    \label{spectral-expansion}
\end{equation}
where $Z$ is the computational domain length in the $z$ direction, corresponding to the longest axial wavelength under consideration. This expansion assumes periodicity with a period of $Z$ in $z$ and ensures analyticity at $r=0$ and harmonic decay at radial infinity, due to the mapped Legendre basis functions $P_{L_n}^m(r)$. In this study, we chose $Z = 20 \pi$, corresponding to the smallest axial wavenumber of $0.1$ to be considered. Meanwhile, $L$, the map parameter, defines the high-resolution region during pseudo-spectral calculations in the range $0 \le r < L$ \citep[see][p. 13]{Lee2023}, and we selected $L=4$ to ensure sufficient resolution for the vortex core and its near periphery.

Although special logarithmic terms may be required to address the $O(1/r)$ decay at large $r$ \citep[see][p. 331]{Matsushima1997}, we omit them here for simplicity in description. The method was originally introduced by \citet{Matsushima1997}, who provided validation examples involving vorticity equations, where further details can be found. An in-depth discussion of the method's implementation in vortex stability research is available in \citet{Lee2023}.

The set of the coefficients $f_{n}^{mk}$ now represents $f$ in a discrete manner. As practical computations require a finite set, we used $n \leq 400$, $|m| \leq 16$, and $|k| \leq 16$. The reason the radial elements make use of an extra degree (i.e., large $n$) compared to the others is to deal with the radially fine structures of the viscous critical layers at high $\Rey$, while a high degree for the other elements (i.e., large $m$ and $k$) is unnecessary since the focus is primarily on small wavenumbers. 

In pure vortex simulations (without particles), the toroidal and poloidal streamfunctions $\psi$ and $\chi$ are discretised in space and then integrated in time to solve the vortex motion governed by \eqref{nonlingoveqn}. To use $\psi$ and $\chi$ as state variables, the toroidal-poloidal decomposition operator $\mathbb{P}$ is applied to both sides of the momentum equation in \eqref{nonlingoveqn}, which leads to the following:
\begin{equation}
    \frac{\partial }{\partial t} \begin{pmatrix}\psi \\ \chi \end{pmatrix} = \mathbb{P} \big( \bm{u} \times \bm{\omega} \big) + \frac{1}{\Rey} {\nabla}^2 \begin{pmatrix}\psi \\ \chi \end{pmatrix}.
\end{equation}
On the other hand, when particles are included, the particle volume fraction $c$ is also considered. The vortex and particle motions are now governed by \eqref{twowaymomeqn}, with $\mathbb{P}$ applied to both sides, i.e.,
\begin{equation}
    \frac{\partial }{\partial t} \begin{pmatrix}\psi \\ \chi \end{pmatrix} = \mathbb{P} \big( \bm{u} \times \bm{\omega} \big) + \frac{1}{\Rey} {\nabla}^2 \begin{pmatrix}\psi \\ \chi \end{pmatrix} - (\vartheta - 1) \,\mathbb{P} \bigg( c \frac{D\bm{u}}{Dt} \bigg),
\end{equation}
for $\psi$ and $\chi$, and \eqref{twowaypvoleqn} for $c$.

When it comes to time integration, the fractional step method is employed, utilising the Adams-Bashforth method for non-stiff terms (e.g., advection) and the Crank-Nicholson method for stiff terms (e.g., viscous dissipation), with Richardson extrapolation applied for the first time step \citep[see][p. 343]{Matsushima1997}. Preliminary simulations with the $q$-vortex ($q=4$) perturbed by a small-amplitude eigenmode with a known frequency and decay rate determined that a time step of $10^{-3}$ yielded a tolerable error for time integration over $t=0$ to $t=100$, corresponding to the time range considered in the main study.

\section{Solenoidal (divergence-free) projection}\label{appC}
For a sufficiently smooth, rapidly decaying three-dimensional vector field $\bm{V}$, the Helmholtz decomposition theorem, with the toroidal-poloidal decomposition as adopted in this study, states that $\bm{V}$ can be expressed in terms of a scalar potential $\theta$ and the toroidal and poloidal scalars $\psi$ and $\chi$:
\begin{equation}
    \bm{V} = - \bm{\nabla}\theta + \bm{\nabla} \times \left\{ \psi \bm{\hat{e}}_z \right\} + \bm{\nabla} \times \left[ \bm{\nabla} \times \left\{ \chi \bm{\hat{e}}_z \right\}\right],
    \label{helmholtz}
\end{equation}
where these three scalar functions are independent of each other and are uniquely determined by imposing the condition that $\theta$, $\psi$, and $\chi$ rapidly decay to zero at infinity. The toroidal-poloidal decomposition operator, $\mathbb{P}$, acts on such vector fields to extract their toroidal and poloidal streamfunctions. Mathematically,
\begin{equation}
\mathbb{P}:\mathcal{U} \longrightarrow \mathcal{P} ~~~\text{such that}~~~ \mathbb{P}(\bm{V}) = (\psi,~ \chi)
\end{equation}
where $\mathcal{U}$ denotes the set of all sufficiently smooth, rapidly decaying three-dimensional vector fields, and $\mathcal{P}$ represents the space of ordered pairs of two sufficiently smooth, rapidly decaying scalar functions. The mathematical details of $\mathbb{P}$ are provided in \citet{Lee2023}, while its computational implementation based on the standard Gauss-Legendre quadrature is explicated in \citet{Matsushima1997}.

On the other hand, given a pair of scalar functions $(\psi, ~\chi)$, the corresponding vector field $\bm{S}$ can be directly constructed as
\begin{equation}
    \bm{\nabla} \times \left\{ \psi \bm{\hat{e}}_z \right\} + \bm{\nabla} \times \left[ \bm{\nabla} \times \left\{ \chi \bm{\hat{e}}_z \right\}\right] = \bm{S}.
    \label{topo}
\end{equation}
We denote this vector field construction as another operator, \(\mathbb{P}^{\dagger}\), defined as
\begin{equation}
\mathbb{P^\dagger}:\mathcal{P} \longrightarrow \mathcal{U} ~~~\text{such that}~~~ \mathbb{P^\dagger}(\psi,~ \chi) = \bm{S}.
\end{equation}
As seen from the difference between \eqref{helmholtz} and \eqref{topo}, $\mathbb{P}^{\dagger} = \mathbb{P}^{-1}$ if and only if we reduce $\mathcal{U}$ to its solenoidal subspace $\mathcal{U}_s \equiv \left\{ \bm{S} ~|~ \bm{S} \in \mathcal{U},~\bm{\nabla}\cdot \bm{S} = 0 \right\}$. In general, \(\mathbb{P}^{\dagger} \big( \mathbb{P} ( \bm{V}) \big)\) can be interpreted as the solenoidal (divergence-free) projection of \(\bm{V}\). As utilised in the present study, the numerical implementation of \(\mathbb{P}\) and \(\mathbb{P}^{\dagger}\) is possible in practice using the mapped Legendre spectral collocation method (i.e., \(\mathsfbi{P}\) and \(\mathsfbi{P}^{\dagger}\) in \S \ref{sec:formulation}).

An important property of $\mathbb{P}$ is that $\mathbb{P}(\bm{\nabla} \theta) = (0,~0)$ for any smooth, rapidly decaying scalar $\theta$ (i.e., if $\bm{\Theta} \equiv \bm{\nabla} \theta$, then $\bm{\Theta}$ has null toroidal and poloidal components). We leverage this property to decouple pressure $p$ (or specific energy $\varphi$) from momentum. For example, if the momentum equation is expressed as $\partial \bm{u}/\partial t = -\bm{\nabla} \varphi + \bm{f} (t)$ under the divergence-free constraint $\bm{\nabla} \cdot \bm{u} = 0$ (where $\bm{f}$ encompasses nonlinear advection, non-conservative forces, etc.), we apply $\mathbb{P}$ to project each term onto the solenoidal subspace. The equation that is actually solved is the projected form, i.e., $\partial \mathbb{P}(\bm{u})/\partial t = \mathbb{P}(\bm{f})$, leading to the following solution form:
\begin{equation}
\bm{u} (t) = \mathbb{P}^{\dagger} \left( \int_0^t \mathbb{P}\big(\bm{f}(\tau)\big) \, d\tau \right) + \bm{u}(0).
\end{equation}
This approach eliminates the typical computational requirement to solve Poisson's equation for \(\varphi\) at every time step to enforce the divergence-free constraint in the momentum equation.

\bibliographystyle{jfm}
\bibliography{jfm_awv_p2}

\end{document}